\documentclass[sigconf]{acmart}

\usepackage{nicefrac}
\usepackage{siunitx}
\usepackage{array,framed}
\usepackage{booktabs}
\usepackage{
  color,
  float,
  epsfig,
  wrapfig,
  graphics,
  graphicx,
  subcaption
}
\usepackage{tikz}
\usepackage{setspace}
\usepackage{latexsym,fancyhdr,url}
\usepackage{enumerate}
\usepackage{algorithm2e}
\usepackage{algpseudocode}
\usepackage{graphics}
\usepackage{xparse} 
\usepackage{xspace}
\usepackage{multirow}
\usepackage{csvsimple}
\usepackage{balance}
\usepackage{amsmath}
\usepackage{caption}
\usepackage[normalem]{ulem}

\usepackage{
  tikz,
  pgfplots,
  pgfplotstable
}
\usepackage{hyperref}

\usetikzlibrary{
  shapes.geometric,
  arrows,
  external,
  pgfplots.groupplots,
  matrix
}
\AtBeginDocument{%
  }

\copyrightyear{2026}
\acmYear{2026}
\setcopyright{cc}
\setcctype{by}
\acmConference[CCS '26]{Proceedings of the 2026 ACM SIGSAC Conference on Computer and Communications Security}{November 15--19, 2026}{The Hague, Netherlands}
\acmBooktitle{Proceedings of the 2026 ACM SIGSAC Conference on Computer and Communications Security (CCS '26), November 15--19, 2026, The Hague, Netherlands}
\acmDOI{10.1145/3830454.3832649}
\acmISBN{979-8-4007-2871-6/2026/11}

\begin{document}

\title{Illusion of Depth: Revealing Hidden Stereo Vision Vulnerabilities in Depth Estimation}


\settopmatter{authorsperrow=4}
\author{Sri Hrushikesh Varma Bhupathiraju}
\affiliation{%
  \institution{University of Florida,}
  \country{Gainesville, USA}
}
\email{bhupathirajus@ufl.edu}
\orcid{0009-0005-1027-5002}

\author{Tetsu Ishizue}
\affiliation{%
  \institution{The University of Electro-Communications,}
  \country{Tokyo, Japan}
}
\email{ishizue@uec.ac.jp}
\orcid{0009-0001-0553-2511}

\author{Nicholas U. Costagliola}
\affiliation{%
  \institution{University of Florida,}
  \country{Gainesville, USA}
}
\email{ncostagliola@ufl.edu}
\orcid{0009-0007-6602-9294}

\author{Ozora Sako}
\affiliation{%
  \institution{Keio University,}
  \country{Tokyo, Japan}
}
\email{sako.ozora@keio.jp}
\orcid{0009-0001-4961-2521}

\author{Kentaro Yoshioka}
\affiliation{%
  \institution{Keio University,}
  \country{Tokyo, Japan}
}
\email{kyoshioka47@keio.jp}
\orcid{0000-0001-5640-2250}

\author{Takeshi Sugawara}
\affiliation{%
  \institution{The University of Electro-Communications,}
  \country{Tokyo, Japan}
}
\email{sugawara@uec.ac.jp}
\orcid{0000-0001-9356-534X}

\author{Sara Rampazzi}
\affiliation{%
  \institution{University of Florida,}
  \country{Gainesville, USA}
}
\email{srampazzi@ufl.edu}
\orcid{0000-0002-3630-6269}

 \renewcommand{\shortauthors}{Sri Hrushikesh Varma Bhupathiraju et al.}


\begin{abstract}
Stereo cameras are integrated into autonomous systems such as self-driving cars, drones, and robots to offer precise depth estimation in a cost-effective manner compared to LiDAR technology. In this work, we reveal an intrinsic vulnerability in stereo cameras that stems from their pixel sampling and calibration processes, which can influence the outputs of stereo matching algorithms. Attackers can achieve fine-grained control over the estimated depth of real obstacles using simple repeating patterns, without relying on sophisticated adversarial machine learning techniques. Furthermore, deep learning-based depth estimation models exhibit a similar vulnerability. We evaluate the impact of this attack on two widely used stereo matching algorithms (BM and SGBM), three deep learning models (PSMNet, MoCha-Stereo, and UniMatch), a stereo-LiDAR fusion model (SGM-DDC), and two popular commercial stereo cameras, the ZED2 and Intel RealSense D435.
For example, in the ZED2 camera, an attacker can displace obstacles up to 20~meters farther or 12~meters closer. In our real-world evaluation in a driving setting, a brief 0.5~second attack can trigger emergency braking in a popular autonomous driving framework. We further demonstrate the feasibility at driving speeds up to 40~km/h using CARLA. Finally, we confirm the ineffectiveness of state-of-the-art defenses, and we propose a novel strategy that leverages similarity scores to dynamically detect and suppress the depth discrepancies. Our work highlights vulnerabilities hidden in stereo matching and deep learning depth estimation models, addressing critical limitations in autonomous system deployments.
\end{abstract}

\ccsdesc[500]{Security and privacy~Hardware attacks and countermeasures}
\ccsdesc[500]{Computer systems organization~Embedded and cyber-physical systems}


\keywords{Cyber-Physical Systems, Physical Attacks, Autonomous Systems, Stereo Cameras}


\maketitle

\begin{figure}[t!]
    \centering
    \includegraphics[width=\linewidth]{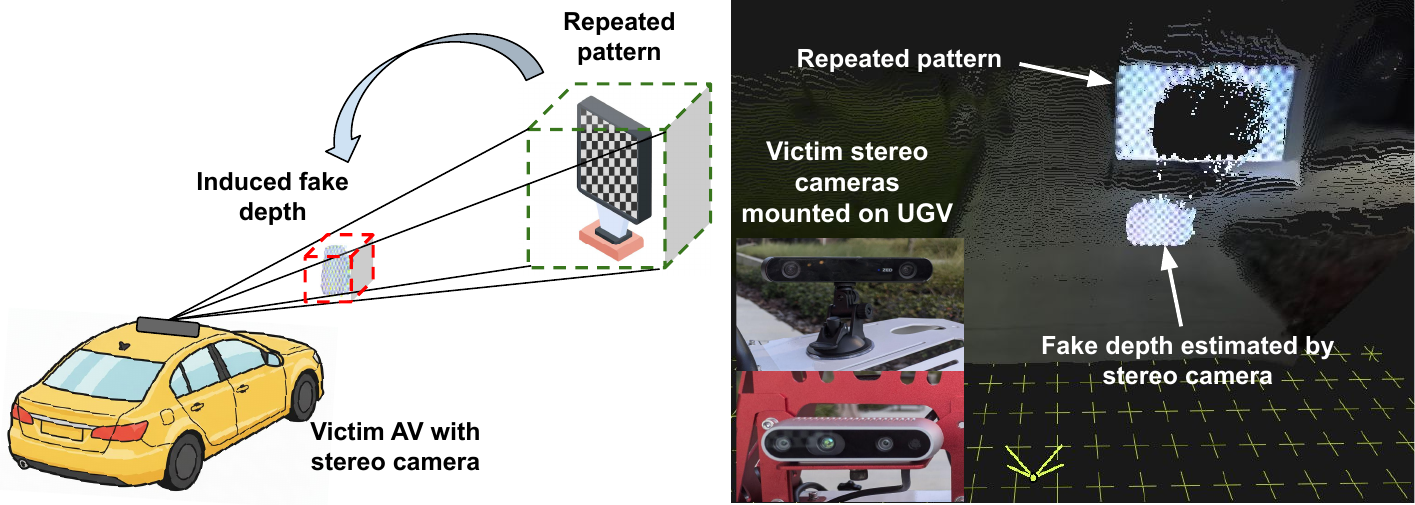}
    \caption{Our attack induces controlled shifts in depth estimation of stereo cameras by exploiting an inherent vulnerability triggered by simple patterns with repeating elements.}
     \Description{Attack overview}
    \label{fig:intro}
\end{figure}

\section{Introduction}

In the evolving landscape of autonomous systems, stereo vision has emerged as a passive, lightweight, and cost-effective alternative solution to LiDARs~\cite{foresight_stereo, nodar}. These sensors mimic human depth perception using dual viewpoints and triangulation, delivering 3D data without sophisticated lasers. For instance, companies such as NODAR~\cite{nodar} and Dolunts~\cite{dolunts} have shown their capabilities in drones and self-driving vehicles under challenging conditions, such as foggy environments, while reducing hardware costs by an order of magnitude~\cite{nodar_blog}. 
At the heart of these systems lie Stereo Depth Estimation (SDE) algorithms and their deep-learning variants, which fuse image data into depth maps of the surrounding environment. Classical matching-based SDE algorithms, such as Block Matching~\cite{geiger2013vision} (BM) and Semi-Global Block Matching~\cite{scharstein2002taxonomy} (SGBM), estimate depth by comparing small \textit{blocks} of pixels between two images. A block from the reference image is horizontally scanned across the other image, and the position with the best similarity is chosen as the match. The horizontal shift, called \textit{disparity}, is then used to compute depth. Deep learning-based SDE models further extend this by learning complex pixel and feature relationships to achieve more accurate and robust disparity predictions~\cite{chang2018pyramid, xu2023unifying}.

The literature has examined adversarial machine learning attacks designed to introduce depth estimation errors in deep learning SDE models~\cite{liu2024physical, chen2024adversary, wang2024left, cheng2022revisiting, wong2021stereopagnosia}. Specifically, recent studies have shown that patterns containing repeated elements, such as stripes, waves, and concentric shapes, can increase attack success rates. These works exploit such periodic structures to generate adversarial examples on benchmark datasets~\cite{liu2025optimization, berger2022stereoscopic}.
However, despite their ability to induce errors, these works neither explain the underlying cause of the phenomenon nor demonstrate control over the resulting depth, restricting their applicability to specific scenarios, complex optimization, and certian deep learning-based SDE models.

This study aims to answer the following critical research questions: \textit{What vulnerability causes depth estimation errors in SDE block matching algorithms and deep-learning models in the presence of repeated elements? Can an adversary exploit such factors to induce consistent and controlled depth estimation manipulation?}

Here, we uncover a hidden vulnerability rooted in two fundamental properties of stereo cameras: \textit{sampling artifacts} and \textit{calibration errors}. Sampling artifacts are pixel-level distortions introduced during image discretization that alter pixel intensities and local features, while calibration errors are pixel-level inaccuracies and nonlinearities caused by lens distortions and manufacturing variations that warp features used for stereo matching.

Our study demonstrates that these vulnerabilities can be exploited to induce incorrect depth estimations, allowing adversaries to systematically bias local matching scores by placing simple patterns with repeating elements in the scene and thereby controlling the resulting depth estimations, as illustrated in Figure~\ref{fig:intro}. These artifacts alter the image pairs at the pixel level, ultimately affecting how SDE algorithms perform their matching.
Moreover, we find that this phenomenon transfers to deep learning-based SDE models: they inadvertently learn and replicate the same vulnerability, making them equally susceptible. Based on this, we design a projection attack that controllably manipulates the depth estimation of both SDE algorithms and deep learning models without any sophisticated adversarial optimization or tracking. 

We systematically formalize and characterize how these vulnerabilities affect depth matching in both classical and deep learning SDE algorithms, with extensive evaluation in diverse scenarios. Specifically, we evaluate the attack on two classical algorithms (BM and SGBM~\cite{scharstein2002taxonomy}), three deep learning models (PSMNet~\cite{chang2018pyramid}, MoCha-Stereo~\cite{chen2024mocha}, and UniMatch~\cite{xu2023unifying}) and one stereo-LiDAR fusion-based model (SGM-DDC~\cite{yao2025stereo}).  
For example, BM can be manipulated to shift obstacles up to 19m closer or 5m farther away from the victim stereo camera, with fine-grained control at 0.1m resolution. Similarly, the attack succeeds on two commercial stereo cameras widely used in autonomous systems (ZED2~\cite{zed2} and RealSense D435~\cite{realsense}),
in realistic driving scenarios, with vehicle speeds up to 15~km/h. We further demonstrate the feasibility of the attack in high-speed scenarios of up to 40~km/h in the CARLA simulator~\cite{dosovitskiy2017carla}. The attack persists for at least 0.5~sec, sufficient to trigger emergency braking or unsafe maneuvers in state-of-the-art autonomous driving frameworks~\cite{autoware}.

While prior work only suggests adversarial training as a potential defense against adversarial examples and impractical manual parameter tuning for reducing depth errors, we propose and evaluate a novel defense technique that detects potentially dangerous repeating patterns in the scene and mitigates the resulting depth shift. The approach analyzes the matching scores of local image regions, suppressing the vulnerability effects.

In summary, our study makes the following contributions:

\begin{itemize}
\item The discovery of a hidden vulnerability in stereo cameras arising from intrinsic characteristics of stereo vision systems, namely (i) sampling artifacts and (ii) calibration errors. The vulnerability can be exploited using simple patterns to control depth estimation on classical block matching and deep learning-based models, which unexpectedly learn and replicate the same vulnerability.
\item An extensive analysis of the attacker capability to manipulate two SDE algorithms (BM and SGBM), three deep learning-based SDE models (PSMNet, MoCha-Stereo, and UniMatch) and a stereo-LiDAR fusion model (SGM-DDC). For example, in synthesized scenarios, an attacker can cause depth estimation errors of up to 58.9m in deep-learning models, producing fake depth estimates as close as 3.7m from the victim camera.
\item Vulnerability evaluation in real-world scenarios of two popular stereo cameras, ZED2 and RealSense D435, achieving depth shifts $>$ 12m. Our attack produces consistent depth estimation errors lasting $\geq$ 0.5~seconds in real‑world driving scenarios at speeds up to 15~km/h, sufficient to trigger emergency braking in autonomous driving systems.\footnote{Details and videos are available at \url{https://cpseclab.github.io/stereocam/}. Artifacts available at \url{https://zenodo.org/records/20767261}. See the Open Policy section for details.}
\item A novel defense methodology that overcomes the limitations of past approaches and detects structured repeated patterns in the scene, mitigating depth estimation errors with a success rate of 96.5\% and 100\% for classical and deep-learning-based SDE algorithms, respectively.
\end{itemize}
\section{Background and Previous Work}
\subsection{Depth Estimation in Stereo Cameras} 
\label{sec:stereo_back}

Stereo-based depth estimation (SDE) reconstructs a 3D depth map by comparing a pair of 2D images captured from two horizontally aligned cameras, mimicking human binocular vision~\cite{szeliski2022computer}. 
Unlike monocular depth estimation systems~\cite{bhoi2019monocular} which infer depth from scale, perspective, or motion, 
SDEs use \textit{disparity}, the horizontal offset between corresponding regions in the left and right images. This disparity is then converted into depth using geometric triangulation. 
SDE algorithms are widely used in autonomous systems~\cite{nodar, honda_work, dji_list} for accurate obstacle detection and avoidance. Our study focuses on the two main categories of SDE: (i) Block matching methods~\cite{scharstein2002taxonomy,lemmens1988survey}, which compute disparity by evaluating local pixel similarity across a predefined window, and (ii) Deep learning-based methods~\cite{chang2018pyramid, xu2023unifying, xu2020aanet}, which leverage high-level semantic features and learned representations via machine-learning algorithms, to establish these correspondences.
While the first category is typically adopted in resource-constrained environments, such as drones and mobile robotics, the second achieves higher accuracy at the cost of increased computational demands, making it more suitable for autonomous driving applications.

\begin{figure}[t]
    \centering
    \includegraphics[width=\linewidth]{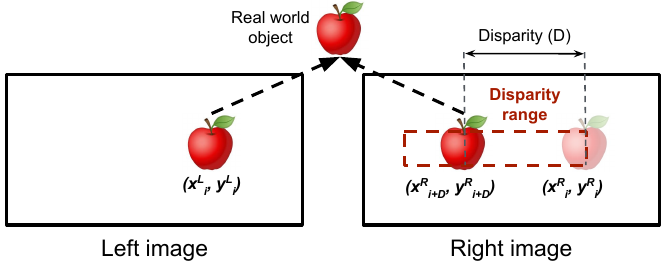}%
    \caption{Depth estimation in stereo cameras. For each block in the left image $(x^L_i,y^L_i)$, classical block matching algorithms start from the same position in the right image $(x^R_i,y^R_i)$ and search for the most similar block $(x^R_{i+D},y^R_{i+D})$ at disparity $D$, within the disparity range.} 
     \Description{Block matching}
    \label{fig:block_mtaching}
\end{figure}

\noindent\textbf{Block matching methods.} In Block Matching (BM) algorithms for stereo cameras, an image (typically the left one) is divided into small rectangular regions, or blocks. For a block in the left image of coordinates $(x^L_i,y^L_i)$, the algorithm examines the same coordinates $(x^R_i,y^R_i)$ in the right image, searching leftward along the same horizontal (epipolar) line for the best block match. The disparity, as illustrated in Figure~\ref{fig:block_mtaching}, is obtained by identifying the most similar block within a defined disparity range. The displacement between the reference block $(x^R_i,y^R_i)$ and the best‑matching block $(x^R_{i+D},y^R_{i+D})$ defines the disparity value $D$, which is subsequently used to estimate the depth. 
The \textit{similarity} between two blocks is typically evaluated using metrics such as the Sum of Absolute Differences (SAD)
or Normalized Cross-Correlation (NCC)~\cite{hirschmuller2008evaluation}.
Semi-Global Block Matching (SGBM)~\cite{lemmens1988survey, scharstein2002taxonomy} methods are enhanced variations of BM that aggregate matching costs along multiple 1D paths. 
BM, SGBM, and their variants are widely used in commercial stereo cameras for drones and robots, such as ZED2~\cite{zed2} and Intel RealSense~\cite{realsense} examined in this work.

\noindent\textbf{Deep learning-based methods.} Deep learning–based SDE models replace the classical matching algorithms, with feature extraction and cost-volume networks that infer disparity end-to-end from the stereo image pairs. Typical model architectures include encoder-decoder or convolutional neural networks (CNNs), and attention mechanisms to give more emphasis to certain correspondences (e.g., edges, textures) and less on ambiguous or noisy regions (e.g., flat surfaces or occlusions)~\cite{hamid2022stereo}. In this work, we analyze three models as representative of three different architectures: PSMNet, MoCha-Stereo, and UniMatch.

\subsection{Attacks against Stereo Depth Estimation}
\label{sec:attack_back}
Researchers have long studied how to induce fake depths to SDEs by generating adversarial examples against their machine learning models. 
Typically, they apply structured pixel-level digital noise or adversarial patches to stereo image pairs to produce erroneous depth measurements~\cite {szegedy2013intriguing, goodfellow2014explaining-old, wong2021stereopagnosia, wang2024left, berger2022stereoscopic}. 
Although these methodologies introduce uncontrolled depth errors, they are highly model-specific, and their applicability in real-world scenarios remains limited because they require direct manipulation of the images or the autonomous system to implement pixel-level modifications.

In contrast, Liu~et~al.~\cite{liu2024physical} created physically realizable adversarial printed patches, leveraging full white-box access to the deep learning (DL) model under attack.
Liu~et~al.~\cite{liu2025optimization} and Berger et al.~\cite{berger2022stereoscopic} instead observed that the presence of repeated elements in the adversarial patches (e.g., stripe, wavy, and concentric designs) increased depth estimation errors. Building on this observation, they generate printed adversarial repeated elements still optimized against specific DL models. Both these works primarily attribute this phenomenon to the susceptibility of the models to adversarially optimized patterns, without examining why such phenomena occur in the first place.
In addition, while all these works succeed on maximizing the depth errors in single stereo image pairs and DL models, they fail to ensure control over the magnitude and direction of the induced errors and their consistency over time and movement, forcing a careful patch crafting, and drastically limiting the attack success in real-world scenarios, like, for example, placing printed posters in front of the car trajectory to succeed.

A separate branch of research focuses instead on stereo camera sensor attacks to induce depth errors. For instance, Zhou~et~al.~\cite{zhou2022doublestar} and Fu~et~al.~\cite{fu2021remote} use light injection to create glare that drones interpret as obstacles. These attacks require one or more light sources to be perfectly aligned with the camera lenses and apply continuous injection to create the glare. While these attacks have been shown to succeed in real-world testing, the success depends on the lens setup of the victim stereo camera, and there is no control over the shape of the glare or the created obstacle. 
Contrary to these works, we demonstrate control over depth estimation of genuine objects under real-world moving conditions, and without the use of adversarial optimization. 
\section{Adversary Model and Vulnerability}

\subsection{Threat Model}
\label{sec:threat_model}

\begin{figure}[t]
    \centering
   \includegraphics[width=\linewidth]{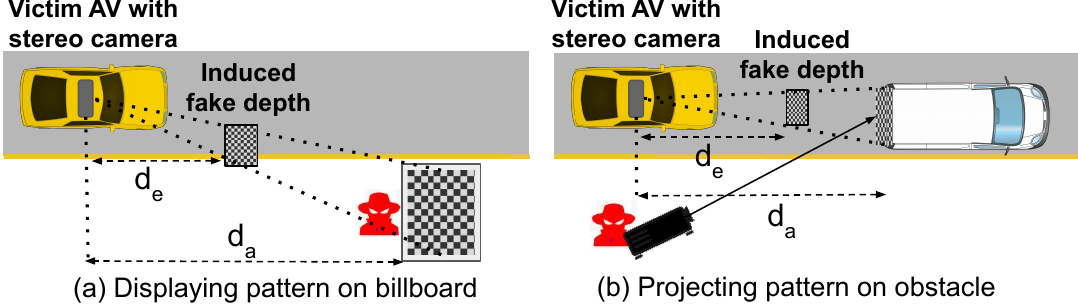}%
    \caption{Illustration of two attack scenarios explored in this work: the projected pattern on a billboard and the back of a van. The attacker controls the fake depth $d_e$ within the frustum of the attack pattern region placed at a distance $d_a$.}
     \Description{Attack scenario}
     \label{fig:attack_scenario}
   \vspace{-3mm}
\end{figure}

\noindent\textbf{Attacker Goal.} We consider stereo cameras deployed in autonomous vehicles (AVs) as a primary target of the attack, as shown in Figure~\ref{fig:attack_scenario}. Different from previous work, which focuses only on maximizing the depth error for a given image pair and deep learning model, here the goal of the attacker is to mislead the victim's AV’s stereo-based perception, either based on classical block matching or deep learning, by consistently controlling the estimated depth of target object surfaces present in the driving scene.  

\noindent\textbf{Underlying Principle and Scenario.}
Instead of generating physical optimized adversarial patches, the adversary can project using a commercial projector simple repeating geometric shapes, such as squares or stripes, onto object surfaces and locations (e.g., buildings, traffic signs, backs of cars or vans) or by displaying them on digital billboards placed at the roadside or intersections (see Figure~\ref{fig:attack_scenario}) for a short time duration (e.g., 0.5 seconds) to activate the attacker's desired outcome. Such scenarios are in line with previous works targeting RGB cameras~\cite{nassi2020phantom, patel2019adaptive, zhou2020deepbillboard}. 
Simply by regulating the geometric structures' shape and size, it is possible to trigger underlying non-linearities and artifacts, forcing disparity mismatch to specific values. In this way, the attacker can manipulate the depth estimation in a controlled and consistent manner without the need for complex depth-specific optimization per scene.
This manipulation causes objects to be perceived as closer or farther than their actual distance from the victim AV. For example, a vehicle may fail to brake and collide by overestimating the distance to an obstacle, whereas underestimating the distance can trigger unnecessary emergency braking or abrupt maneuvers, compromising user safety.

\noindent\textbf{Assumptions.} We assume that the adversary has knowledge of the type of stereo camera used by the victim AV and the type of the SDE algorithm (i.e., BM or deep learning-based model), the corresponding algorithmic parameters (e.g, disparity range, block size, stabilization, and confidence thresholds), and the calibration settings. This information can be inferred from publicly available camera manuals and data sheets, like the ones analyzed in this work~\cite{zed2, realsense}. While proprietary algorithms are not fully disclosed, stereocamera vendors publish technical specifications (window size, disparity range, etc.) to comply with standards and interoperability requirements. Alternatively, the attacker can perform a black-box study by acquiring a stereo camera similar to the one used in the victim AV. Based on this assumption, we perform the attack on two widely used commercial stereo cameras, ZED2~\cite{zed2} and RealSense D435~\cite{realsense}, both of which employ proprietary SDE algorithms.

The attack is remote, not requiring any firmware or hardware access to the victim vehicle or stereo camera. We do assume oracle access to the depth estimation algorithm employed by the victim stereo camera, consistent with the assumptions from previous works~\cite{liu2025optimization, berger2022stereoscopic}. Finally, we focus on easy, realizable, low-effort repeated geometric shapes (e.g., simple stripes or chessboard patterns) that an unskilled attacker can project over flat or semi-flat surfaces without requiring sophisticated optimization, white-box model access, or the generation of adversarial examples. An adversary can select more sophisticated geometries to account for the specific structure of the object surface to project on.

First we synthesize such scenarios on the KITTI dataset~\cite{geiger2013vision} and evaluate classical and deep learning-based SDE models in Section~\ref{sec:evaluation}. Real-world dynamic evaluation in indoor and outdoor settings are presented in Section~\ref{sec:real_world_eval}, using commercial RealSense and ZED2 stereo cameras.

\subsection{Vulnerability in Block Matching Algorithms}
\label{sec:attack_principles}

\begin{figure}[t]
    \centering
   \includegraphics[width=\linewidth]{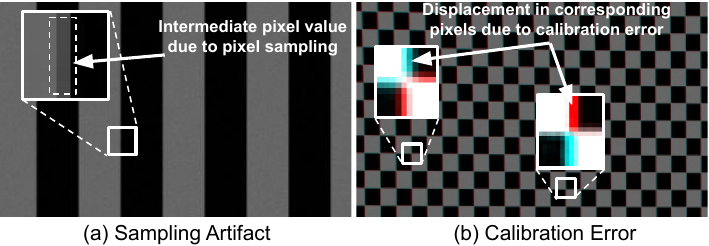}%
    \caption{(a) Sampling artifacts produce intermediate pixel intensity values at the stripe boundaries due to averaging across the pixel grid in our Gazebo simulation\protect\footnotemark. (b) A red-cyan anaglyph of the stereo pair reveals calibration-induced geometric distortion: the cyan (left image) and red (right image) misalignments cause spatially varying shifts in disparity across a checkerboard pattern.} 
     \Description{Sampling artifact visualization}
     \label{fig:stereo_causalities}
\end{figure}

\footnotetext{Due to grayscale conversion in Gazebo visualization rendering, the pattern appears as gray and black stripes instead of black and white.}

This section describes the principles that enables attackers to precisely manipulate perceived depth by exploiting (i) sampling artifacts and (ii) calibration errors in stereo cameras, which are empirically validated on the BM and the ZED2 SGBM algorithms.

\begin{figure}[t]
    \centering
   \includegraphics[width=\linewidth]{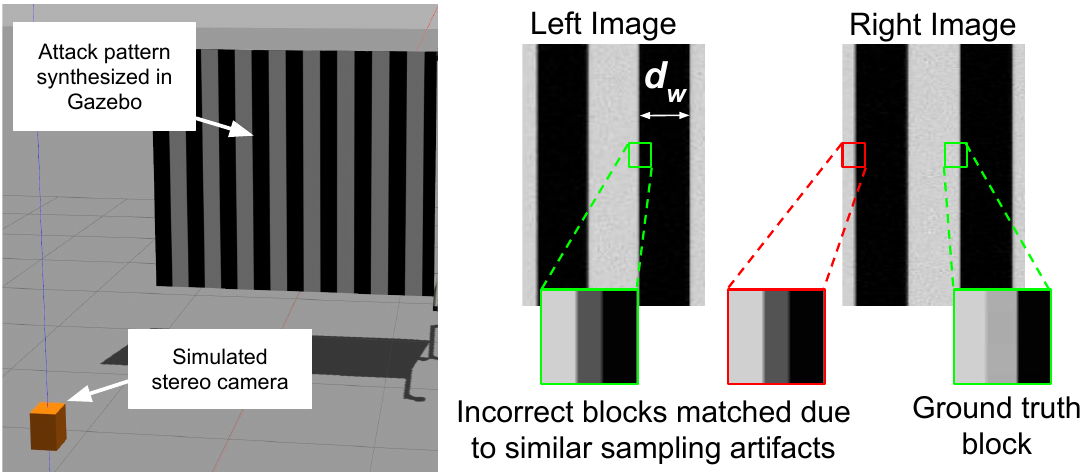}%
    \caption{Gazebo simulation scenario with stripe pattern in front of stereo camera (left). Illustration of the incorrect block matching triggered by sampling artifacts (right).} 
     \Description{Gazebo setup}
    \label{fig:sampling_feasibility_result}
\end{figure}

\subsubsection{Sampling Artifacts}
\label{sec:sampling_principle}

Image sensors in stereo cameras capture continuous physical scenes and represent them with discrete pixels~\cite{lorre1980artifacts}. 
The light intensity recorded at each pixel represents the average optical light intensity from that region of the scene.

Given a physical point $p = (x, y, z)$ 
and its optical light intensity $I_\mathrm{opt}(p)$, the corresponding pixel intensity value $P_i$ is denoted by
\begin{equation}
P_i = \frac{1}{A(SR_i)} \int_{SR_i}^{SR_{i+1}} I_\mathrm{opt}(p) \, \mathrm{d}A
\hspace{1em} 
\label{eq:pixel_intensity1}
\end{equation}
where $SR_i$ is the sampling range of the $i^{th}$ pixel in the horizontal axis, defined as the minimum real-world distance represented by a single pixel~\footnote{For this formalization, we restrict our analysis to the horizontal axis, consistent with the matching direction employed in classical block matching algorithms~\cite{sgbm_opencv}}.
Assuming square pixels, the physical area $A$ corresponding to the $i^{th}$ pixel is given by $A(SR_i) = SR_i^2 $.

When the stereo camera captures images with high‑contrast boundaries (e.g., object edges) in the physical space, each image sensor captures and discretizes the boundary, resulting in subtle pixel-level inconsistencies. The pixel intensity values in those particular regions consequently assumes an intermediate value corresponding to the average intensity of the region sampled by the stereo camera, as illustrated in Figure~\ref{fig:stereo_causalities} (a). These intermediate values, referred to as \textit{sampling artifacts}, are inherent to imaging processes~\cite{lorre1980artifacts} and can produce a measurable disparity offset in SDE algorithms and, consequently, changes in the estimated depth.

\noindent\textbf{Formalization.} In scenes containing repeated high‑contrast boundaries, such as a black-white chessboard pattern, the sampling artifacts exhibit differences in both pixel intensity and spatial position. Although the physical pattern does not change, each repetition of the geometric shape is captured differently depending on the boundary's alignment with the left and right cameras.

To examine this effect, we employ a classical BM algorithm that uses the SAD score, as described in Section~\ref{sec:stereo_back}, to estimate disparities. The SAD score is computed by summing the absolute differences between corresponding pixel intensities in two image blocks ($B(i)$), providing a measure of their similarity.
We synthesize a simple stripe pattern in the popular Gazebo simulation environment~\cite{Koenig2004Gazebo} and capture stereo images using an emulated ZED2 stereo camera, as shown in Figure~\ref{fig:sampling_feasibility_result} (left). 

Figure~\ref{fig:sampling_feasibility_result} (right) illustrates the left and right images of the stripe pattern synthesized in Gazebo. $B(i_{\mathrm{Ref}})$ denotes the reference block in the left image, and $B(i_T)$ the true corresponding block in the right image, both containing the artifacts (gray pixels). The red block marks the region with a matching sampling artifact that is incorrectly paired with the left image.

\begin{figure}[t]
    \centering
   \includegraphics[width=\linewidth]{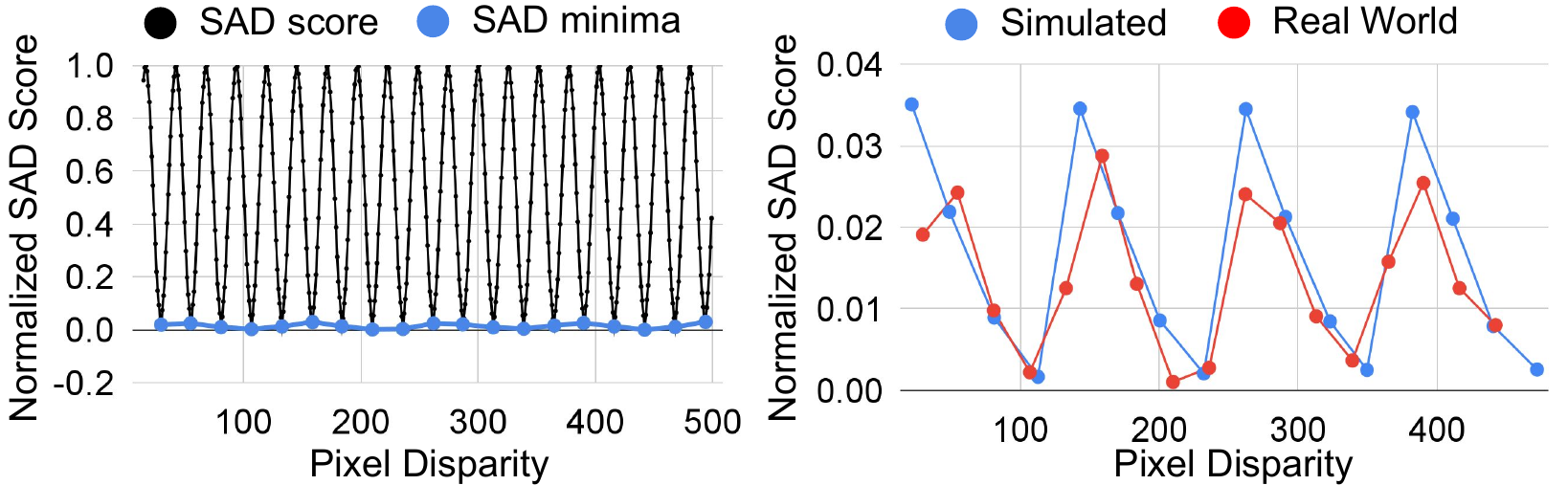}%
    \caption{Normalized SAD scores following an oscillating trend due to the black and white stripe pattern (left). The SAD minima in simulated and real-world scenarios (right). The normalized SAD scores are scaled in the (right) graph to better illustrate the SAD minima.} 
     \Description{Sampling artifact causality}
    \label{fig:sad_minima_variance}
\end{figure}

The BM algorithm selects the global minimum in the SAD score as the optimal match. Under natural conditions, the algorithm converges at the minimum matching score of $\mathrm{SAD}(B(i_{\mathrm{Ref}}), B(i_T))$, finding the perfect match. However, the high-contrast alternate pattern and resulting artifacts produce oscillating SAD scores, resulting in multiple SAD minima, each of which is a potential match, as shown in Figure~\ref{fig:sad_minima_variance} (left). Let $SA(i)$ denote the sampling artifact magnitude at the block position $i$, estimated from the pixel intensity. Based on this, the SAD score at the minima is proportional to the difference in pixel intensities, described as:

\begin{equation}
\mathrm{SAD}(B(i_{\mathrm{Ref}}), B(i_T+n\cdot d_w)) \propto \left|SA(i_{\mathrm{Ref}})-SA(i_T+n\cdot d_w)\right|
\end{equation}

Where $n \cdot d_w$ represents a horizontal shift by $n$ repetitions of the pattern, with $n \in \{1, 2, \cdots\}$. Furthermore, each repeated element in the pattern yields a similar SAD score because it reappears every $d_w$ pixels.
Thus, the BM algorithm selects the repeated block whose sampling artifact magnitude (and consequently SAD minima) most closely matches that of the reference block, rather than necessarily the true geometric match, producing an incorrect disparity estimate. Appendix~\ref{appendix:sad_influence} provides a detailed formulation of the vulnerability. Since $SA(i)$ varies periodically with the repeating pattern, an adversary can exploit this predictable behavior to manipulate the location of the SAD minima and systematically bias stereo depth estimation. The adversary can simply craft repeated geometric elements in the pattern whose distances (e.g., adjusting $d_w$) produce the sampling artifacts necessary to shift the estimated depth to the desired value and trigger controlled matching of selected regions.

\noindent\textbf{Real‑World Validation.}
We further validate these findings using real‑world images captured with a ZED2 stereo camera, with the repeated stripe pattern displayed on a monitor, placed at $d_a=1$m in controlled indoor conditions ($d_w$ = 12mm as in the simulation). Despite the pixel-level noise in the stereo image pairs, the minima in the normalized SAD scores vary according to the sampling artifact magnitude, consistent with the simulation results, as illustrated in Figure~\ref{fig:sad_minima_variance} (right). This experiment confirms that sampling artifacts can be used to control disparity estimation.

\subsubsection{Calibration Error}
\label{sec:calibration_principle}
SDE algorithms use calibrated image pairs for disparity estimation~\cite{szeliski2022computer}. Stereo camera vendors typically perform a factory calibration process to determine the parameters to pre-process the raw camera images. This critical process establishes a precise mapping between a 3D point in the physical world and its corresponding 2D projection in the image, compensating for imperfections such as lens aberration or misalignment between the lens and the sensor plane~\cite{abraham2005fish}. 
However, unavoidable nonlinearities cause pixel correspondence errors that persist in the reconstructed depth map~\cite{xiong1997error}.
These distortions arise because corresponding pixels in the left and right images are displaced in different directions, resulting in mismatches along both the horizontal and vertical axes, as shown in Figure~\ref{fig:stereo_causalities} (b). Although such calibration errors are inherent to stereo systems, they usually have little impact in benign cases. Yet, in the presence of high-contrast boundaries such as the ones present in repeated patterns, the calibration errors warp across each repeated element, causing the correct match to appear structurally different in the image pairs, and consequently trigger an incorrect match. The resulting mapping depends on the camera's parameters (such as focal length, principal point, and translational and rotational matrices) and magnitude of the calibration error.

\noindent\textbf{Formalization.} To examine how calibration errors cause depth estimation inaccuracies, we model the calibration error of our ZED2 camera using the fundamental parameterization of radial and tangential distortions~\cite{beauchemin2001modelling}, and the calibration models provided by the ZED2~\cite{zed2} vendor.
The parameterization is as follows:

\begin{equation}
\label{eq:right_camera_distortion}
\begin{aligned}
\begin{pmatrix} x_d \\ y_d \end{pmatrix} 
&= \begin{pmatrix} x_u \\ y_u \end{pmatrix}
\left( 1 + k_1 r^2 + k_2 r^4 + k_3 r^6 \right) \\
&\quad + 
\begin{pmatrix} 
2 a_1 x_u y_u + a_2 (r^2 + 2 x_u^2) \\ 
a_1 (r^2 + 2 y_u^2) + 2 a_2 x_u y_u 
\end{pmatrix}
\end{aligned}
\end{equation}

Here, $(x_d, y_d)$ and $(x_u, y_u)$ denote the pixel coordinates of the distorted and undistorted images, respectively. The distortion vector $C = (k_1, k_2, a_1, a_2, k_3)^T$ comprises the radial distortion parameters ($k_1, k_2, k_3$) and tangential distortion parameters ($a_1, a_2$), and $r$ represents the radial distance of the image pixel. 

Using the standard stereo calibration technique~\cite{abraham2005fish}, we estimate the distortion parameters for the ZED2 cameras, which are then applied to Equation~\ref{eq:right_camera_distortion}. Using this formulation, we quantify the residual calibration error that remains in the camera after factory calibration and rectification.
The resulting distortion vectors are $C_L$ = (-0.0415, 0.0099, 0.00035, -0.00013, -0.00492)$^T$ and $C_R$ = (-0.0424, 0.0112, 0.00015, 9.69$\times10^{-5}$, -0.00509)$^T$,
where $C_L$ and $C_R$ are the displacement vectors for the left and right cameras, respectively. 
This calibration error causes pixel shift as shown in Figure~\ref{fig:stereo_causalities} (b).
The calibration error magnitude ($E_C$) is estimated 3.7~pixels in the ZED2 camera, derived by the maximum distance between points $(x_d, y_d)$ and $(x_u, y_u)$ in Equation~\ref{eq:right_camera_distortion}. This parameterization captures semantic‑level calibration errors but cannot account for pixel‑level calibration inaccuracies. Calibration is based on finite mathematical approximations that inherently fail to model stochastic noise, non‑planar surface imperfections, and higher‑order lens irregularities. A 3~pixel calibration error is generally considered acceptable for a commercial stereo camera~\cite{abraham2005fish}. Nevertheless, this level of error is sufficient to cause depth estimation errors in the presence of repeated elements.

To evaluate the effect of calibration error on BM, we use the same Gazebo simulation setup of the sampling artifact simulation. In this case, we employ a chessboard pattern, since calibration errors occur along both the horizontal and vertical axes. We define \textit{calibration error magnitude} ($E_C$) as the largest pixel shift induced by calibration error, ranging from 0-15 pixels at intervals of 3 (up to five times the estimated calibration error magnitude of ZED2, 3.7 pixels). This range in calibration error is used to describe the relationship between calibration error and depth estimation. 

\begin{figure}[t]
    \centering
   \includegraphics[width=\linewidth]{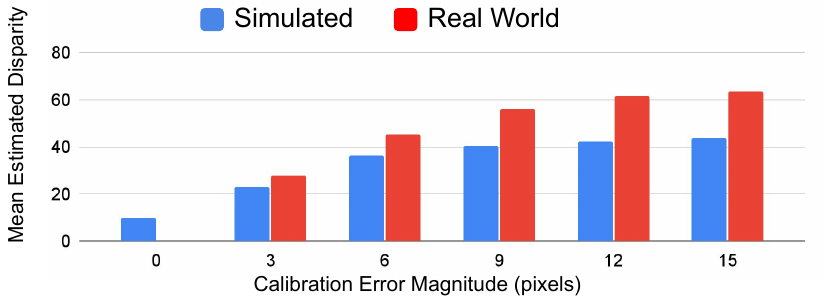}%
    \caption{Estimated disparity variation at increasing calibration error in simulated and real-world conditions. Note that ZED2 has a calibration error of 3 pixels by design.} 
     \Description{Disparity variation}
    \label{fig:calibration_feasibility_result}
   \vspace{-3mm}
\end{figure}

Consistent with the methodology in Section~\ref{sec:sampling_principle}, we set $d_w$= 12mm to ensure the pattern is perceptible to the ZED2 camera.
Unlike sampling artifacts, which influence the minima of the SAD scores, calibration errors affect the global score matching of the BM algorithm. For this reason, we consider the mean of the estimated disparity within the chessboard pattern as our metric. The estimated mean disparity values at increasing calibration error magnitudes are shown in Figure~\ref{fig:calibration_feasibility_result}. As hypothesized, even at $E_C =$ 3 pixels the estimated mean disparity doubles, increasing from 9.9 at zero error to 22.8 in our simulation. The disparity continues to rise until an error magnitude of 12, after which it plateaus. This plateau occurs because the BM algorithm’s disparity range (default value of 64) caps the estimation, preventing further increase and consequently flattening the average value.
An adversary can select a pattern with $d_w$ such that the perceived pixel size of its geometric shapes is close to $E_C$ and increase the estimated disparity. These results demonstrate how calibration errors can be exploited to induce incorrect block matching.

\noindent\textbf{Real-World Validation.}
We validate our simulation results with real‑world experiments, using the same setup as in Section~\ref{sec:sampling_principle}. The chessboard pattern displayed on a monitor at $d_a$=1m is captured with the ZED2 camera, starting from $E_C=$ 3 up to 15 pixels. 
As illustrated in Figure~\ref{fig:calibration_feasibility_result}, we observe a similar trend with the mean disparity increases until magnitude 12 and then plateaus following the disparity range value of the ZED2 SGBM algorithm. Unlike simulation, the mean disparity values reach higher values (up to 63), indicating that pixel imperfections in real stereo images further amplify the effect.

\subsection{Vulnerability in Deep Learning Models}
\label{sec:DL-characterization}

Deep learning-based SDE models employ convolutional, encoder-decoder, or attention-based architectures to learn and predict correspondences between high-level features extracted from the left and right images, as described in Section~\ref{sec:stereo_back}. In contrast with classical block matching algorithms, they are generally trained to estimate the matching cost for each pixel in the full disparity range, followed by cost aggregation techniques that integrate local and global information to suppress noise in the disparity distribution~\cite{liu2025optimization, chang2018pyramid, xu2020aanet}.

Here, we experimentally verify the presence and effect of the discovered vulnerability in one of the most popular depth estimation models used in autonomous vehicles, PSMNet~\cite{chang2018pyramid} (full characterization is detailed in Section \ref{sec:stereo_back}). To achieve this, we train the model using the standard autonomous driving KITTI~\cite{geiger2013vision} dataset under default parameters.
By following the Gazebo formalization described in Section~\ref{sec:attack_principles}, we synthesize stripe and chessboard patterns with sampling and calibration artifacts. 
The synthesized patterns are then embedded into 100 randomly chosen stereo image pairs from the KITTI dataset, emulating a pattern projected on the back surface of a van or other obstacle, 10m in front of the victim vehicle stereo camera (equivalent to $\approx$ 18\% of the image pairs with a true disparity of 31).
The patterns are luminance-matched for each KITTI image by computing the image’s average brightness over the whole scene (calculated as luma~\cite{ituR_Luma}) 
and shifting the pattern’s average luminance to match the KITTI scenes.

\subsubsection{Sampling Artifacts} 
\label{sec:sampling_deeplearning}

\begin{figure}[t]
    \centering
   \includegraphics[width=\linewidth]{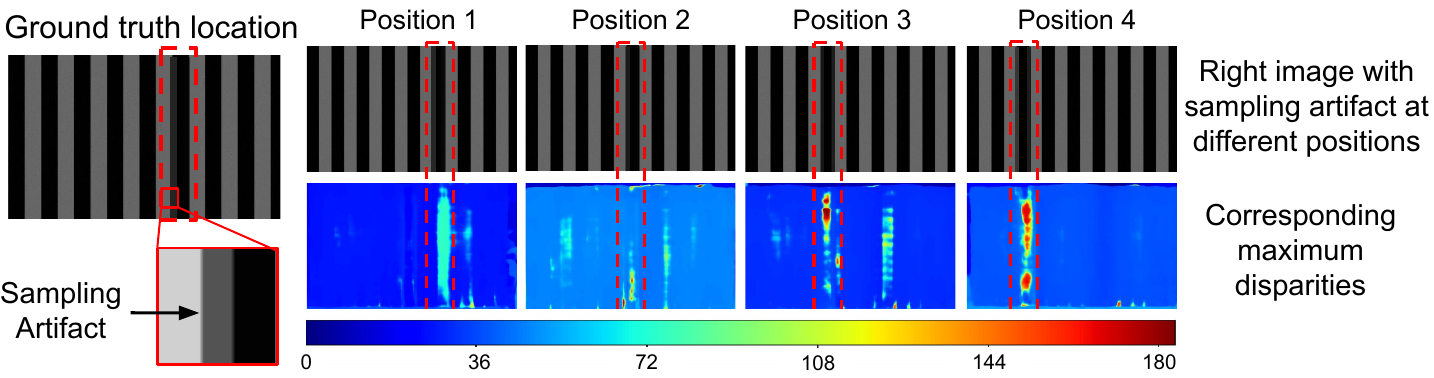}%
    \caption{The synthesized patterns with sampling artifacts at different positions and the corresponding depth maps from PSMNet. The change in the maximum disparity corresponds to the sampling artifact position.} 
     \Description{Sampling artifact ablation}
     \label{fig:calibration_dl_feasibility_setup}
\end{figure}

To demonstrate if sampling artifacts affect the depth estimation in PSMNet, we introduce the artifacts at selected stripes within the pattern, beginning from the ground truth location (Position 1), as shown in Figure~\ref{fig:calibration_dl_feasibility_setup}. 
The sampling artifact magnitude (pixel intensity) in the left image at Position 1 is fixed, while in the right image, the artifact is iteratively shifted in the next stripe to the left.
This shift causes the corresponding maximum disparity to increase following the artifact position shift, as seen in the corresponding depth maps.

\begin{figure}[t]
    \centering
   \includegraphics[width=\linewidth]{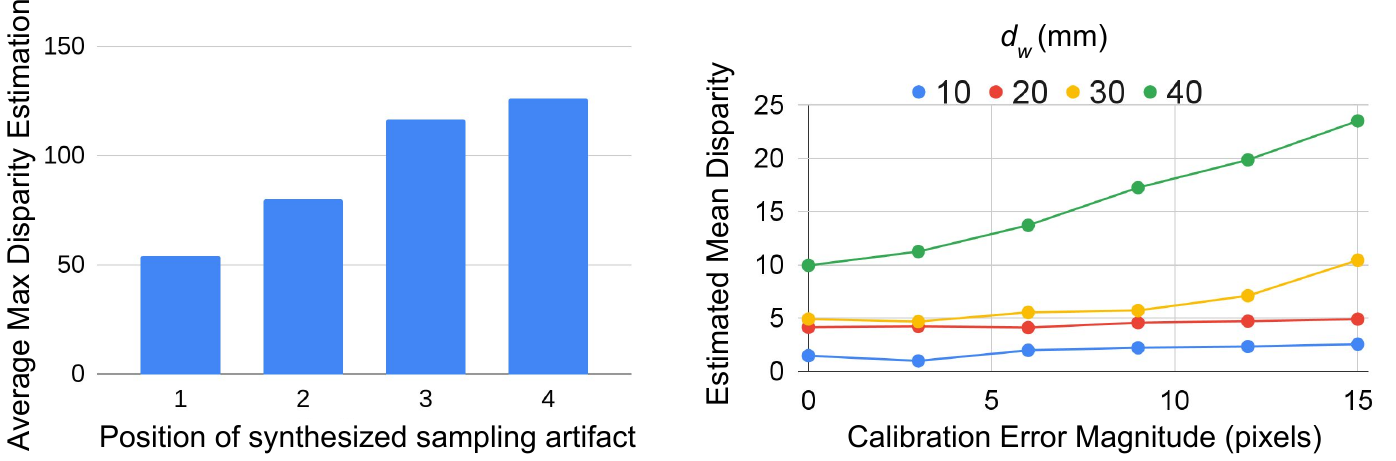}%
    \caption{The average maximum disparity estimation at increasing positions of the synthesized artifact (left). The mean disparity estimation for PSMNet at calibration error magnitudes at increasing $d_w$ (right). The graphs show that feature matching in deep learning-based SDE models is influenced by sampling artifacts and calibration errors.} 
     \Description{Calibration ablation}
     \label{fig:calibration_dl_feasibility_result}
\end{figure}

Figure~\ref{fig:calibration_dl_feasibility_result} (left) shows that the average maximum predicted disparity increases over the 100 images, as the sampling artifact is shifted farther from the ground‑truth position (from Position 1 to 4). 
These results demonstrate that the feature correspondences learned by the ML models are directly influenced by the location and magnitude of the sampling artifacts. Consequently, incorrect depth is estimated whenever a sampling artifact of similar magnitude appears. We further validate the influence of the sampling artifact pixel intensity on the disparity estimation of PSMNet in Appendix~\ref{appendix:sampling_consistency}.

\subsubsection{Calibration Error} 
\label{sec:calibration_deeplearning}

As with the sampling artifacts, we conduct a similar evaluation to demonstrate that calibration errors also influence the depth estimation in PSMNet.
We consider a chessboard pattern with increasing chessboard‑block widths $d_w$ from 10 to 40mm. 
We then apply calibration errors ranging from 0 to 15 pixels to 100 randomly selected KITTI images. 

The resulting disparity estimations produced by PSMNet are shown in Figure~\ref{fig:calibration_dl_feasibility_result} (right). The estimated disparity increases as the calibration‑error magnitude grows, similar to the behavior observed for the BM algorithm in Section~\ref{sec:calibration_principle}.

At larger $d_w$ (above 40mm), the effect of calibration on the structure of the chessboard pattern is reduced, limiting its influence on disparity estimation. This is consistent with the BM algorithm behavior and confirms that the vulnerability observed in the BM and SGBM algorithms also extends to PSMNet.

\section{Attack Characterization}
\label{sec:characterization}

As defined in the threat model, the attack consists of displaying or projecting on a flat or semi-flat target surface with a simple repeated pattern crafted to trigger the desired sampling artifacts and calibration errors, leading to shift in depth.
This section analyzes the stereo matching process and the attacker capabilities with respect to the following parameters: (i) Pattern design, (ii) Algorithmic block matching parameters (block size and search range), (iii) Distance from the victim stereo camera, (iv) Physical size of the pattern, and (v) Contrast of the pattern. An attacker can systematically adjust the above parameters to achieve the desired depth shift.

For this study, all simulations are conducted in Gazebo, where a virtual stereo camera is configured to match the ZED2 camera specifications by the vendor. We then validate the analysis in controlled real-world indoor scenarios, projecting the patterns using a ViewSonic PA700W projector~\cite{viewsonic}. In these real-world experiments, we collect 10 images of each pattern and average them to minimize the pixel noise in the image pairs. For characterization, we consider the BM algorithm with the SAD score as our reference metric.

\begin{figure}[t]
    \centering
   \includegraphics[width=\linewidth]{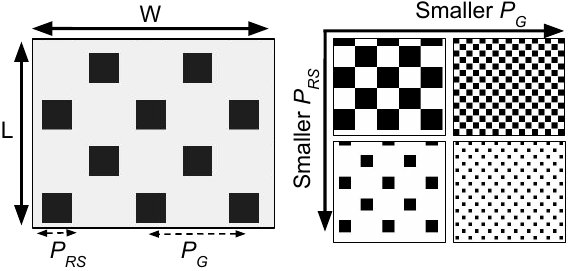}%
    \caption{Parameterization of the pattern based on granularity of the chessboard pattern ($P_G$), and size of the repeating black squares in the pattern ($P_{RS}$).} 
     \Description{Pattern structure}
     \label{fig:pattern_parameterization}
\end{figure}

\noindent\textbf{Formalization.} An adversary places a structured attack pattern of physical width $W$ and length $L$, at a distance $d_a$ from the victim stereo camera, as illustrated in Figure~\ref{fig:intro}. The relative pixel size of the pattern $(w_p,l_p)$ in the image pair can be measured as
\begin{equation}
\begin{bmatrix}
w_p \\
l_p
\end{bmatrix}
=
\frac{1}{d_a}
\begin{bmatrix}
f_x & 0 \\
0 & f_y
\end{bmatrix}
\begin{bmatrix}
W \\
L
\end{bmatrix}
\hspace{1em} 
\label{eq:pixel_to_mm}
\end{equation}
where $f_x$ and $f_y$ denote the focal lengths of the camera in the x and y-directions, respectively. The objective is to induce a desired depth estimation value, defined as $d_e$, by exploiting the sampling artifacts and calibration error. To achieve this, we parameterize a chessboard pattern, representing the simplest form of geometric shape repetition. 
The structure is characterized by \textit{granularity} ($P_G$), which defines the width of repetition, and the \textit{repeating element size} ($P_{RS}$), as illustrated in Figure~\ref{fig:pattern_parameterization}. The adversary can measure the $d_w$ corresponding to desired $P_G$ and $P_{RS}$, using Equation~\ref{eq:pixel_to_mm}, by setting $W = L = d_w$.

\subsection{Pattern Design}
\label{sec:pattern_design}
To achieve control over depth estimation, an attacker varies $P_G$ and $P_{RS}$ to assess the relationship between the pattern and the corresponding disparity measurement. We span the granularity $P_G$ from 1 to 50 pixels, beyond the typical block size used by SGBM/BM algorithms (i.e., 15–30) and the size of the chessboard black square $P_{RS} \in [1, P_G]$ for each $P_G$. 
For example, we set $d_a$ to 1.5m and the size of the chessboard pattern in Gazebo to 1.5$\times$1m ($W \times L$ $\approx$ 490 $\times$ 300 pixels), which emulates the rear surface of a typical car obstacle and within a standard size of a roadside billboard~\cite{patel2021overriding}. The BM algorithm is applied with a block size of 15 pixels and a disparity range of 64 (default). We then collect the relative change in the disparity achieved by each pattern. The attack is considered successful in changing the measured disparity if more than 10 pixels exhibit a disparity change, consistent with thresholds in popular autonomous driving frameworks for obstacle detection (e.g., Autoware~\cite{autoware}).
The correct disparity in the given scenario is 21 pixels (relative disparity 0 with no attack). 

To confirm the consistency of our results also in real-world setting, we project the same patterns in front of the ZED2 camera, setting the initial granularity to $P_G=3$ pixels (35mm), given the resolution of the ZED2.

\begin{figure}[t]
    \centering
   \includegraphics[width=\linewidth]{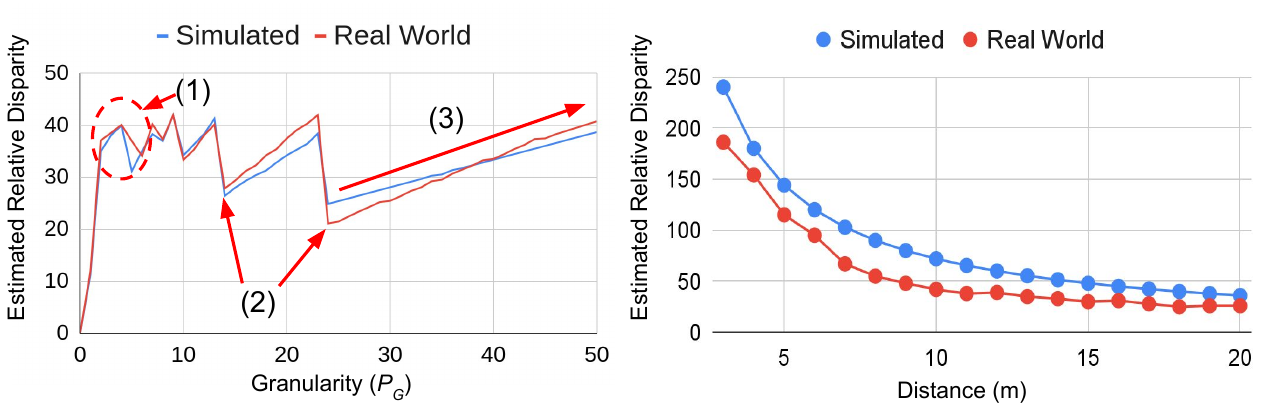}%
    \caption{Relative disparity errors in BM with respect to pattern design and contrast (left). The trend shows (1) a mismatch due to calibration error, (2) periodic drops in disparity depending on the position and magnitude of the closest sampling artifact, and (3) a steady, linear increase in disparity error.
    Disparity error at increasing distances (right).} 
     \Description{Characterization results}
    \label{fig:design_capability}
\end{figure}

\noindent\textbf{Results.} The trend in the estimated relative disparity as $P_G$ increases in the BM algorithm for both simulation and real-world evaluation is illustrated in Figure~\ref{fig:design_capability} (left). In both cases, the disparity exhibits a cyclic pattern, increasing with $P_G$ until the disparity range value is reached, followed by a sharp drop and a subsequent repetition of the upward trend. These abrupt drops (e.g., toward disparity values of 60–70) occur because increasing $P_G$ alters the sampling artifact magnitude, shifting it away from the reference block and outside the disparity range associated with the current false match. As a result, the matching cost reaches a new temporary minimum, leading the algorithm to select a different correspondence based on the similarity caused by the sampling artifacts. 
The granularity $P_G$ can be adjusted to exploit the calibration error or to control specific depth estimations by shifting the sampling artifact position. For example, within the $P_G$ ranges of 14–23 pixels (46-76mm) or 23–50 pixels (76-165mm), a linear increase in estimated disparity can be leveraged for control. The analysis also shows that changes in $P_{RS}$ do not affect the BM algorithm's results, since BM matches blocks strictly horizontally. In Section~\ref{sec:evaluation}, we show that SGBM and deep learning–based SDE models, which rely on pixels and features across all directions, are influenced by the $P_{RS}$ value.

\noindent\textbf{Domination of Calibration Error.}
At lower granularity (up to 6 pixels $P_G$), the calibration error dominates the effect of the sampling artifact, producing a large depth shift that reaches the disparity range limit of the BM algorithm (64). This occurs because, although sampling artifacts are formed on the image, they are displaced and warped by geometric distortions, altering the artifact pixels in a manner consistent with the modeled pattern structure (see Equation~\ref{eq:right_camera_distortion}). Given a calibration error magnitude $E_C$ and a sampling range $SR$, the projected pattern width $d_w$ must satisfy $d_w \geq E_C \cdot SR$ in order to exceed the calibration error.

\noindent\textbf{Block Matching Algorithm Parameters.} 
Maintaining the same $d_a$, $P_G$, and pattern size, we investigate how BM algorithm parameters, such as block size and disparity range, influence the disparity estimation. We increase the block size from 5 to 105 at intervals of 10 pixels (since blocks in BM algorithm need to have a well-defined center, they cannot be divisible by 2), until the stereo matching algorithm cannot predict any relevant depth information. We then increase the disparity range from 16 to 160 (disparity range divisible by 16 is a standard practice~\cite{sgbm_opencv}). 

We observe that increasing the block size expands the trend horizontally, as a larger pattern granularity ($P_G$) is required to induce disparity errors. This reflects the fact that with larger blocks, the algorithm aggregates more pixel information, thereby shifting the point at which repeated structures begin to cause mismatches. In contrast, the disparity range expands the trend vertically by limiting the range of possible disparity values. While the vulnerability triggered by the chessboard pattern remains unchanged, the amplitude of the cyclic disparity errors is capped by this disparity range, as illustrated in Figure~\ref{fig:design_capability}. Thus, the size of the block determines the threshold required to manipulate the disparity, whereas the range of the disparity limits the magnitude of the estimated disparity while keeping the influence of granularity unchanged.

\subsection {Attack Design}
\label{sec:attack_scenario}

In this section, we examine the basic geometry underlying depth estimation errors and analyze how the distance and the spatial region of the repeated pattern influence the resulting depth estimates.

In stereo vision, the cameras observe the scene through a viewing volume known as the \textit{camera frustum}, which defines the spatial region within which objects can be captured, and depth can be estimated. Figure~\ref{fig:attack_scenario} illustrates the typical frustum of a stereo camera, consisting of a truncated pyramid extending from the camera center, with its width and height at a given depth $Z$ determined by the field of view and focal length $f$~\cite{ccoltekin2006foveation}. For an object with physical size $S$ at an original depth $d_a$, its projected size on the image plane ($S_{\text{P}}$) scales inversely with depth according to $S_{\text{P}} = S \cdot d_a / d_e$, where $d_e$ is the induced depth after manipulating disparity, as illustrated in Figure~\ref{fig:attack_scenario}. 
When the target object depth is altered to be perceived closer to the stereo camera (decreasing $d_e$), $S_{\text{P}}$ increases, and as it is moved farther, $S_{\text{P}}$ decreases, reflecting the perspective scaling within the frustum. All depth manipulations conducted by the attacker are therefore constrained to this frustum volume, with the size of the object varying according to its position within the frustum. Measuring the frustum, it is possible to characterize how the induced disparity changes with pattern distances and sizes.

\subsubsection{Distance from the Stereo Camera.} 
\label{sec:pattern_distance}
Using a pattern design with $P_G$ and $P_{RS}$ of 23 pixels (maximum disparity error from experiments in Section~\ref{sec:pattern_design}), we vary the distance between the stereo camera and the projection surface, $d_a$, from 3-20m in 1m increments, in the Gazebo environment. The projected pattern has a physical width of 1.5m (240 pixels in image space). As $d_a$ increases, the perceived size of the pattern decreases proportionally to $f/d_a$, where $f$ is the focal length of the camera. The disparity estimated by the BM algorithm exhibits a similar decreasing trend with increasing $d_a$, as shown in Figure~\ref{fig:design_capability} (right) in both simulated and real-world testing with the projected pattern. This behavior occurs because the number of potential matching candidates for stereo correspondence decreases as the projected pattern occupies fewer pixels in the images, thereby reducing the maximum achievable disparity error. 
We further show in Section~\ref{sec:indoor_dist_robustness} that this trend persists also in RealSense D435.

\subsubsection{Physical Size of the Pattern.} To analyze the relationship between pattern size and disparity error, we vary the width ($W$) and length ($L$) of the projected pattern from 0.5 to 3~meters in increments of 0.5~meters, while fixing the projection distance at $d_a=3$~meters and keeping the pattern design constant. Both Gazebo-based simulations and real-world experiments show that the estimated disparity is directly influenced by the width of the pattern in pixel space: as the physical size of the pattern increases, its width in pixels also increases, leading to a corresponding rise in disparity estimation. This increase follows an approximately linear trend, starting from 44 pixels for $W$ = 0.5m, until the disparity reaches the disparity range. Beyond this point, the error saturates and remains fixed at the disparity range value.
This behavior arises because stereo matching algorithms search for correspondences only within a predefined disparity range; once this range is exceeded, additional pattern size no longer reduces disparity error. These results indicate that, depending on the stereo depth estimation algorithm implemented in a target stereo camera, an adversary could estimate this limit and design pattern sizes that reliably induce erroneous depth perception. Further, based on the camera frustum formalization described earlier in this section, if an attack region of width $W$ is required at distance $d_a$, then the width $W'$ needed at a different attack distance $d'_a$ can be estimated as $W' = W \cdot d'_a/d_a$. Similarly, for a given attack region of width $W$, the corresponding distance $d_a$ required to achieve the attack can be obtained from the inverse relation. Based on these properties of disparity estimation, an adversary can design the attack using the relation $d_a = \frac{f \cdot W}{D}$, where $f$ is the camera’s focal length and $D$ is the estimated disparity. Here, $d_a$ indicates the distance to the pattern and $W$ its physical width. To induce the same depth estimation error at a greater distance, the attacker can simply scale the pattern by this relation.

\subsubsection{Pattern Contrast.} 
\label{sec:capability_contrast}
Sampling artifacts typically arise at boundaries between different pixels, where their intensity takes on an intermediate value between the intensities on either side. Even when the contrast in the pattern is small, these artifacts can still influence stereo matching by altering the perceived pixel structures. To demonstrate this, we select the pattern design corresponding to the maximum disparity error in Figure~\ref{fig:design_capability}, and progressively reduce its contrast in both the Gazebo simulation and real-world scenario by 10\% increments ($\approx$25 pixel intensity levels) up to 90\%, as a 100\% reduction would remove the pattern. 
We find that the disparity error persists until the contrast is reduced to 80\%, in both simulation and real-world conditions. Below this level, the magnitude of the sampling artifact becomes too small to cause disparity errors. 

Based on the above characterization, the adversary can analytically derive the pattern design for a target $d_e$, camera distance $d_a$, pattern size and contrast, without optimizing for a specific scene.

\section{Evaluation on Driving Dataset }
\label{sec:evaluation}

Using the characterization outlined in Section~\ref{sec:characterization}, we perform a comprehensive analysis of attack effectiveness and induced depth control for both BM and SGBM algorithms implemented via the OpenCV library~\cite{sgbm_opencv}, deep-learning-based SDE models (PSMNet~\cite{chang2018pyramid}, MoCha-Stereo~\cite{chen2024mocha} and UniMatch~\cite{xu2023unifying}), and stereo-LiDAR fusion-based model (SGM-DDC~\cite{yao2025stereo}) over the entire KITTI driving dataset. 

\subsection{Experimental Setup}
\label{sec:experiment_setup}

For this evaluation, we collect real-world chessboard patterns displayed on a monitor placed at a distance of $d_a=1$m to account for sampling artifacts and calibration errors in the captured stereo pairs. We then synthesize these patterns using the methodology described in Section~\ref{sec:DL-characterization} to overlay them onto all 7518 stereo image pairs from the KITTI dataset.  Figure~\ref{fig:kitti_synthesis} in the Appendix illustrates an example synthesized image.

The patterns are placed in front of the victim stereo camera, with a fixed true discrepancy ranging between 5 and 35~pixels, representing an equivalent real-world physical distance between 62 and 9m, emulating potential vehicles and billboard distances and consistent with assumptions adopted in prior work~\cite{liu2025optimization, liu2024physical}. We evaluate 627 patterns across 7 true discrepancies, resulting in 4389 total combinations per stereo pair. 
We extract the relative disparity change and assume the attack is successful only when a minimum of a 10-pixel region exhibits the desired disparity. This is consistent with the 10‑point threshold used for obstacle detection in state-of-the-art autonomous driving frameworks such as Baidu Apollo and Autoware~\cite{apollo, autoware}. Note that for this evaluation, the calibration error magnitude in the synthesized pattern is 3~pixels, as measured in ZED2. To ensure visibility of the pattern, we set $P_G$ and $P_{RS}$ initial values to 3~pixels, as finer granularity cannot be resolved by our ZED2 stereo camera due to its resolution.

\subsection{Stereo Matching Algorithms}
\label{sec:classifc_eval}

For both BM and SGBM algorithms, we set the block size to 15~pixels and the disparity search range to 64~pixels, following the default configurations of the algorithms under analysis~\cite{sgbm_opencv}. 
In BM, we observe that the maximum induced depth exhibits a similar trend, consistent with the pattern shown in Figure~\ref{fig:design_capability}. In BM, a $P_{RS}>5$ pixels is necessary to shift the predicted depth, since smaller values result in patterns that are too fine to be reliably captured by the victim stereo camera. 
Within a $P_G$ range of 23–63 pixels, which provides precise control over disparity estimation, the attack achieves fine-grained manipulation at the single-pixel disparity level. This translates to control over the fake‑depth distance $d_e$, allowing an adversary to move the pattern surface from 1m to 13m in front of the camera.

In contrast, SGBM shows a different behavior, where the maximum disparity increases with simultaneous growth in $P_G$ and $P_{RS}$. This occurs because SGBM introduces smoothness constraints by aggregating costs across multiple directions. While this multi-path aggregation suppresses outliers and reduces vulnerability to locally ambiguous matches, it makes SGBM less sensitive to lower $P_G$ values but susceptible to higher ones. Similar to BM, the attack demonstrates a fine-grained control over $d_e$ between 11.5m and 2.5m (disparities 26-58). Furthermore, a larger $P_{RS}$ increases the vertical similarity considered by the SGBM algorithm, resulting in a higher calculated disparity. 

\noindent\textbf{Natural Motion.}
\label{sec:blurriness}
In natural driving scenarios, the camera’s finite exposure duration produces motion‑blur artifacts in regions where scene elements move relative to the camera. This blur smears object boundaries and alters pixel intensities through sampling artifacts and region‑specific calibration errors. To evaluate the attack under realistic motion, we synthesize motion blur on KITTI images using the experimental setup described in Section~\ref{sec:experiment_setup}. We test the attack on the BM and SGBM algorithms. Motion blur is generated using the point spread function~\cite{szeliski2022computer}. The details of the motion blur simulation are provided in Appendix~\ref{appendix:motionblur}. We apply the point spread function assuming a vehicle moving at a constant speed of 40~km/h, resulting in an average displacement of approximately four pixels for the projected pattern.

We observe that the attack exhibits the same overall behavior for BM and SGBM, consistent with Section~\ref{sec:classifc_eval}. For BM, the results follow the trend shown in Figure~\ref{fig:design_capability}, indicating precise control of depth estimation within a $P_G$ range of 23–63~pixels, with a maximum $d_e$ of 12m. For SGBM, the maximum disparity increases jointly with $P_G$ and $P_{RS}$, yielding a maximum $d_e$ of 9m. This aligns with Section~\ref{sec:classifc_eval}, where multi‑path aggregation in SGBM suppresses larger depth errors. Overall, these results show that the sampling artifacts and calibration errors introduced by the patterns persist and enable controlled depth manipulation despite motion‑blur at speeds of 40~km/h.

\subsection{Deep Learning SDE algorithms}
\label{sec:DL-eval}
Here, we evaluate the shift in estimated depth on the state-of-the-art PSMNet~\cite{chang2018pyramid}, MoCha-Stereo~\cite{chen2024mocha} and UniMatch~\cite{xu2023unifying} models, trained on the KITTI dataset~\cite{szeliski2022computer}, using their default training parameters.  
Specifically, we consider the median calculated disparity of 80\% of the pattern region (from the center), and measure the shift in desired depth due to $P_G$ and $P_{RS}$ variation. Table~\ref{tab:deep_learning_results} lists the maximum disparity estimation error induced by the attack at increasing ground truth disparities. 

\begin{figure*}[t]
    \centering
   \includegraphics[width=\linewidth]{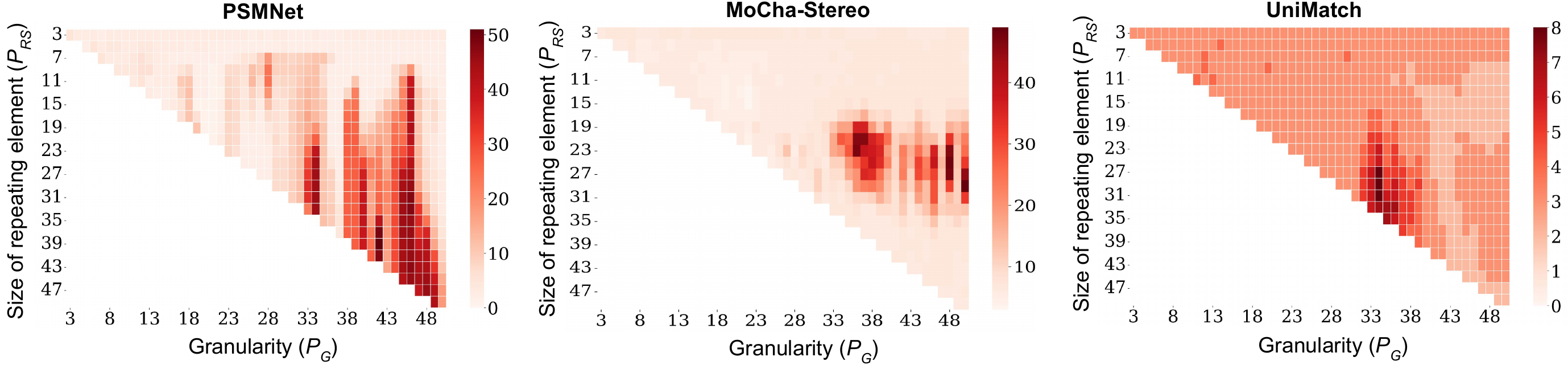}%
    \caption{Example of heatmaps showing the maximum predicted disparity for PSMNet, Mocha, and UniMatch models, reported as a delta relative to the ground-truth disparity.} 
     \Description{Deep learning results}
    \label{fig:deltas}
\end{figure*}

\noindent\textbf{PSMNet Analysis.}
In PSMNet, the attacker can increase the calculated disparity by up to 79 pixels relative to a pattern placed at a 5 pixel disparity (ground truth). This corresponds to a 59m shift in estimated depth toward the victim camera from a pattern located at $d_a$=62m. Similar to BM and SGBM, $P_{RS}$ > 5 pixels is necessary to induce a substantial shift in the calculated disparity, as the block must be large enough for the model to reliably detect and match. We observe a cyclic change in estimated depth with increasing $P_G$ of the pattern geometry, as seen in Figure~\ref{fig:deltas}. Fine‑grained control over the estimated depth can be achieved with an average step size of approximately 1 pixel. 

As the true disparity increases, the achievable range of depth shift decreases as expected. 
Additionally, we find that the calculated disparity can increase up to 22 pixels, corresponding to a controllable increase in depth estimation of up to 15m.

\begin{table}[t]
\centering
\caption{Maximum disparity estimation error induced by the attack across the tested ground truth disparities.}
\label{tab:deep_learning_results}

\scalebox{1.0}{
\begin{tabular}{c|ccccccc}
\hline
\multirow{3}{*}{\textbf{Model}} &
\multicolumn{7}{c}{\textbf{Ground Truth Disparity (pixels)}} \\
\cline{2-8}
& \small{\textbf{5}} & \small{\textbf{10}} & \small{\textbf{15}} & \small{\textbf{20}} & \small{\textbf{25}} & \small{\textbf{30}} & \small{\textbf{35}} \\
\hline

PSMNet & 79 & 77 & 51 & 55 & 51 & 72 & 49 \\
MoCha-Stereo & 30 & 69 & 34 & 70 & 51 & 69 & 73 \\
UniMatch & 187 & 132 & 12 & 10 & 8 & 15 & 20 \\

\hline
\end{tabular}}
\end{table}

\noindent\textbf{MoCha-Stereo Analysis.} 
The attack causes a controlled depth estimation error of up to 52m, corresponding to an estimated depth of 10m in front of the victim camera at a true distance of 62m. We observe a similar cyclic change in estimated depth with increasing $P_G$ as seen in PSMNet, with a higher sensitivity to $P_{RS}$ variation (Figure~\ref{fig:deltas}). Control over the estimated depth is achieved with a step size between 0.9 and 1.55 pixels, consistent across all disparities. 

\noindent\textbf{UniMatch Analysis.} In UniMatch, we observe a shift in calculated disparity of up to 187 pixels, equivalent to an estimated depth of 60.5m towards the victim camera at a true distance of 62m. In contrast with PSMNet, UniMatch presents a higher effect with a smaller $P_{RS}$. We observe similar peak granularities to PSMNet at select $P_G$ (e.g., 33-34) as shown in Figure \ref{fig:deltas}, indicating transferability of selected patterns between PSMNet and UniMatch. We also achieve controllable depth estimation between 1 and 187 pixels, with an interval between 0.5 and 1.2 pixels, except at a true disparity of 10, where a single outlier increases the interval to 19 pixels. Furthermore, we find that calculated disparity can increase up to 20 pixels, corresponding to an increase in depth estimation of 12m.

In contrast with PSMNet, between 10 and 20m, we achieve a maximum depth control of just 6 pixels ($\sim$2.4m).

\subsection{SDE Stereo - LiDAR Fusion Model}
\label{sec:fusion_eval}

We further extend our evaluation to a Stereo-LiDAR Fusion Model, SGM-DDC~\cite{yao2025stereo}. This model augments the semi-global matching (SGM) algorithm with a discrete disparity-matching cost (DDC) to insert sparse LiDAR disparities into the SGM cost volume. First, sparse LiDAR disparities are propagated to neighboring pixels to form a semi-dense LiDAR prior. The SGM cost volume is then compared against this LiDAR prior, and the DDC term favors disparities where both the stereocamera and LiDAR measures agree with each other. A final consistency check is used to ensure cohesion between the sensor views. 

To evaluate the fusion model against our patterns, we synthesize a LiDAR point cloud to represent a realistic flat wall, such as a billboard displaying the chessboard pattern.

The results show a maximum change in depth from 62m to 16m from the target vehicle, a shift in depth of 46m.
This estimated depth error decreases quickly within 31m, with a shift of just 4.4m possible with a more restricted range of control, varying from 4 and 11 pixels depending on the true depth of the pattern, with an interval between 0.64 and 1.43 pixels. When the pattern is placed between 63m and 21m away from the camera, the range of control in depth is only 4 pixels. 
This reduction in range of control and overall error corresponds with a reduced effect of $P_G$ variation, particularly at farther distances. SGM-DDC explicitly favors disparities closer to the LiDAR prior, which is unaffected by the pattern $P_G$ and $P_{RS}$. This constrains the attacker to induce a shift in depth estimation consistent with the LiDAR prior. SGM-DDC is therefore more resilient to pattern variation than the non-fusion models, but it remains affected by repeated patterns.

\section{Real World Evaluation}
\label{sec:real_world_eval}

We examine the effectiveness of the attack in static indoor scenarios and dynamic outdoor settings with the ZED2~\cite{zed2} and Intel RealSense D435~\cite{realsense} stereo cameras, using the ViewSonic PA700W to project the patterns over a wall and the back of a van. 
To assess end‑to‑end consequences on the autonomous driving framework we use the Euclidean clustering within Autoware~\cite{autoware} to perform obstacle detection, and consider the attack successful if the false depth is detected as an obstacle.

\subsection{Static Indoor Scenario}
\label{sec:indoor_basic}
\begin{figure}[t]
    \centering
   \includegraphics[width=\linewidth]{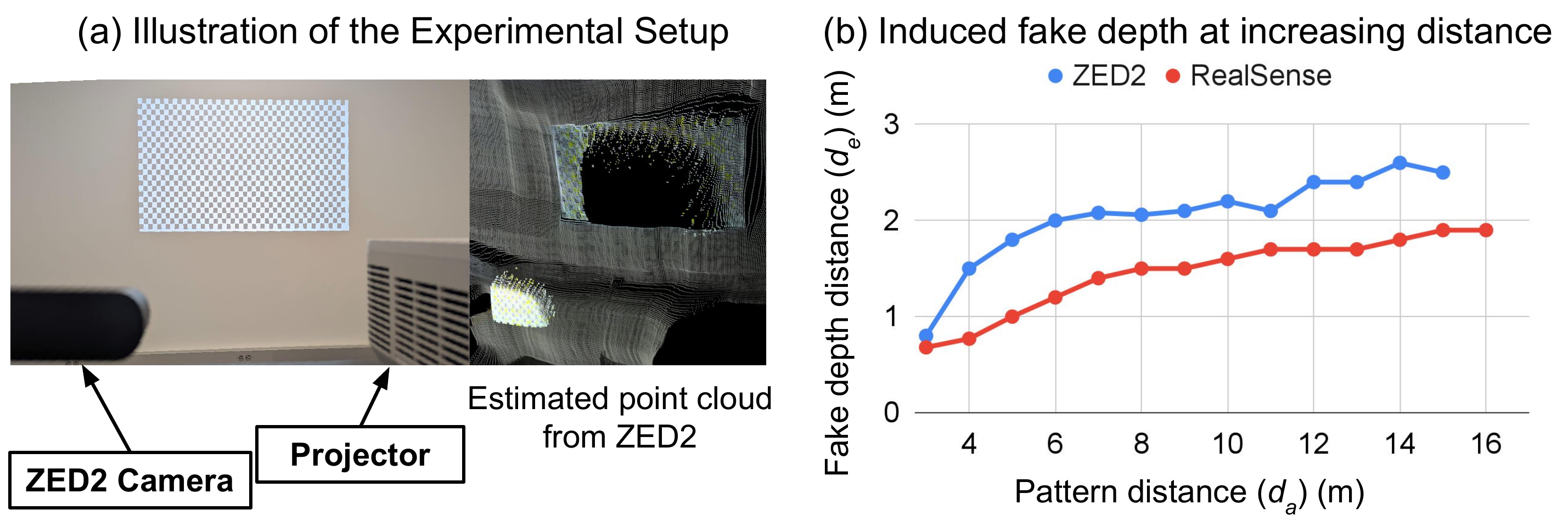}%
    \caption{(a) Indoor scenario setup. (b) Induced depth error ($d_e$) at increasing pattern distances ($d_a$) for ZED2 and RealSense cameras in indoor scenario .} 
     \Description{Indoor setup and results}
     \label{fig:distance_robustness}
\end{figure}

To analyze the shift in estimated depth and controllability of the attack under real-world conditions, we set $d_a$ = 3m, with the chessboard patterns projected onto a flat surface, consistent with Section~\ref{sec:attack_scenario}. The distance of 3m is chosen to ensure that the entire projected pattern is captured within both the left and right images of the tested stereo cameras, and the ambient illumination is maintained at a constant value of 500~lux (typical of an indoor laboratory environment). At $d_a$ = 3m, a minimum pattern width $W$ of 0.9m and length $L$ of 0.6cm is required to induce false depth in ZED2, while $W$ of 60~cm suffices for RealSense. These widths correspond to $\leq$ 1.4\% of the image pair area in both ZED2 and RealSense cameras. The attack produces a shift in depth of up to 2.4m in RealSense and 2.2m in ZED2, meaning the surface is perceived at 0.6m and 0.8m distance to the camera, as illustrated in Figure~\ref{fig:distance_robustness} (a).

Following the characterization of the stereo matching algorithms, we evaluate the controllability of the attack on both cameras using chessboard patterns with increasing granularity ($P_G$). Please note that we apply this for ZED2 and RealSense without reverse engineering or knowledge of their proprietary SGBM, using the same pattern against classical BMs.

\noindent\textbf{Results and Observations.} For the RealSense camera, the pattern enables fine-grained control, with depth adjustable in increments of 0.1m between 0.4 and 2.6m. The depth increases linearly as $P_G$ varies from 14 to 27 pixels. Since the RealSense camera has a depth range of 3m, the attack does not displace the pattern farther from its original distance. In contrast, the ZED2 camera exhibits coarser control, with changes in depth of approximately 2.2, 1.4, and 0.5m observed at $P_G$ = 27, 33, and 46~pixels, respectively. At lower values of $P_G$ (5-10), the attack shifts the estimated depth farther by up to 4m, while maintaining fine-grained control at approximately 1m resolution. Moreover, the RealSense camera requires a repeating geometric shape size to satisfy $P_{RS} = P_G$ (a complete chessboard pattern) to induce false depth, whereas the ZED2 camera requires $P_{RS}$ such that $P_G - 7 < P_{RS} < P_G$. We hypothesize that this difference arises because ZED2 employs a more advanced stereo matching algorithm that incorporates features such as edges and contours, reducing fine-grained control over depth estimation. Nonetheless, the evaluation demonstrates that it is possible to induce consistent false depth as close as 0.8m from the victim camera in both ZED2 and RealSense. Note that, for each experiment, a depth shift is considered successful if the resulting point‑cloud points are detected as a genuine obstacle by Autoware’s Euclidean clustering.

\subsubsection{Controllability at Increasing Distances} 
\label{sec:indoor_dist_robustness}
The distance between the stereo camera and the projection surface is increased from $d_a = 3$m to 20m at increments of 1m.  Building on the characterization of the attack in Section~\ref{sec:pattern_distance}, we project a pattern of $W \times L$ = $2 \times 1.5$m, as the prior results demonstrate that this dimension is necessary to induce false depth at extended distances. We select pattern with $P_G$ and $P_{RS}$ that resulted in the highest depth shifts for ZED2 and RealSense, respectively. For each distance, the experiment is repeated 10 times.

Figure~\ref{fig:distance_robustness} (b) presents the minimum induced false depth ($d_e$) in RealSense and ZED2 stereo cameras at increasing projection distances. 
The attack can shift depth estimation from up to $d_a$= 15m in ZED2 and 16m in RealSense. For example, the attack can shift the surface depth as close as 2m from both cameras, when $d_a = 10$m.
The trend in $d_e$ initially increases exponentially and then transitions to a near-linear saturation when $d_a > 12$m for RealSense and $d_a$ > 8m for ZED2. This behavior is consistent with the experimental results in Figure~\ref{fig:design_capability} (right), where the incorrect disparity values decrease exponentially at first and then approach saturation at larger $d_a$. These findings demonstrate the controllability of the attack from distances of up to 15m.

\subsubsection{Scene Illumination}
\label{sec:robust_lighting}
To evaluate the attack under varying illumination, we use the same experimental setup as in Section~\ref{sec:indoor_basic} with the patterns corresponding to highest shift in depth. The ambient illumination in the room increases from 0 to 1000~lux, in steps of 100~lux. For each illumination level, the evaluation is repeated 10 times, and the estimated depth maps from the ZED2 and RealSense cameras are collected. Across all the tested illumination conditions, the estimated false depth remains consistently at 0.8m for ZED2 and 0.6m for RealSense. These findings align with the results in Section~\ref{sec:capability_contrast}, confirming that sampling artifacts affect stereo matching even under higher illumination conditions.

\subsection{Outdoor Evaluation}
\label{sec:static_outdoor}
For our outdoor realistic scenarios, we collect estimated depth maps from the ZED2 and RealSense cameras mounted on an AgileX Hunter 2.0~\cite{agilex} unmanned ground vehicle (UGV). The surface is positioned 1.5m to the right of the UGV, emulating a roadside billboard as shown in Figure~\ref{fig:attack_scenario}, with $W \times L$ = $2 \times 1.5$m. 

Under nighttime conditions with an ambient illumination of 50~lux, the attack produces similar trends to the indoor setting. For example, the estimated depth plateau at 2.2m for ZED2 and 1.8m for RealSense when $d_a$ = 10m. Similarly, the attack induces false depth up to $d_a$ = 16m for ZED2 and 18m for RealSense. The slightly extended attack range observed at night is attributed to lower ambient illumination, which enhances the projected pattern sharpness and contrast. 

Under daytime conditions, with ambient illumination ranging from 1000 to 1500~lux, the attack induces false depth, causing obstacle detection at $d_e$ = 2.5m in front of the camera when $d_a$ = 8m for ZED2, and at $d_e$ = 1.8m, when $d_a$ = 10m for RealSense. The reduced attack range is due to the higher illumination, which diminishes the brightness of the projected pattern captured by the sensor, consistent with the contrast results of Section~\ref{sec:capability_contrast}. This reduction in brightness lowers the effective contrast of the pattern and decreases the magnitude of sampling artifacts. Nevertheless, these results show the practicality of the attack in real-world conditions. 

Using the same experimental setup, we collect stereo image pairs with the victim cameras mounted on a UGV traveling at 10~km/h (refer to Section~\ref{sec:ethics} for details), at distances to the projected pattern ranging from $d_a$ = 20 to 3m. We consider the attack successful, if consistent depth shift for $\geq$ 0.5~sec duration is achieved, sufficient to trigger emergency braking in AD systems such as Autoware~\cite{autoware}.

The consistent attack durations achieved in these experiments are listed in Table~\ref{tab:attack_duration}. In nighttime conditions, the attack with $P_G$ = 49 and $P_{RS}$ = 39 produces a stable false depth of 2.8m in ZED2 for 4.6~sec and approximately 1.3m in RealSense for 3.2~sec. Under daytime illumination, the attack with $P_G$ = 30 and $P_{RS}$ = 25 achieves consistent false depth for up to 2.1~sec in ZED2 and 1.8~sec in RealSense. These results are similar to those in static scenarios with a 0.3m variance. Furthermore, the attack requires higher $P_G$ and $P_{RS}$ than in the indoor experiments in Section~\ref{sec:indoor_basic}, since a successful attack requires consistent depth shift from further distances. These results show the possibility for attackers to consistently induce false depth for a sufficient amount of time to trigger automatic responses in the autonomous system, such as unexpected and sudden braking or dangerous maneuvers.

\begin{figure}[t]
    \centering
   \includegraphics[width=\linewidth]{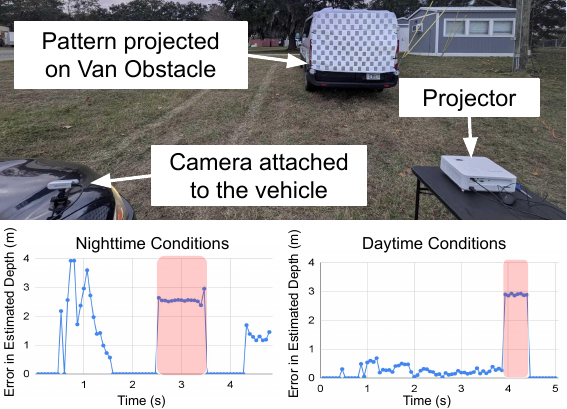}%
    \caption{Experimental setup in outdoor scenarios, with van obstacle (top). The induced depth by the ZED2 camera in night (bottom-left) and daytime (bottom-right) conditions.} 
    \Description{Outdoor setup and results.}
    \label{fig:van_experiment_setup}
   \vspace{-3mm}
\end{figure}

\begin{table}[t]
\centering
\caption{Attack duration (in seconds) for both RealSense D435 and ZED2 cameras in night and daytime conditions.}
\label{tab:attack_duration}
\scalebox{0.9}{

\begin{tabular}{c|cc|cc}
\hline
\multirow{2}{*}{\textbf{Condition}} &
\multicolumn{2}{c|}{\textbf{10 km/h (UGV)}} &
\multicolumn{2}{c}{\textbf{15 km/h (car)}} \\
&
\textbf{RealSense D435} & \textbf{ZED2} &
\textbf{RealSense D435} & \textbf{ZED2} \\
\hline

Daytime   & 1.8s & 2.1s & 1.0s & 0.7s \\
Nighttime & 3.2s & 4.6s & 1.2s & 1.0s \\

\hline
\end{tabular}}
\end{table}

\subsubsection{Driving Scenario} 
\label{sec:realworld_scenario}
We extend the evaluation to a real-world scenario where an attacker projects a pattern onto a van obstacle positioned either on the roadside or in an adjacent lane to the victim vehicle, as illustrated in Figure~\ref{fig:attack_scenario}. The projector, placed 4m behind the van, casts a 1.8m wide attack pattern onto its rear surface as shown in Figure~\ref{fig:van_experiment_setup}. The van is located 1.5m to the side of the victim car, and the depth maps of the ZED2 and RealSense are collected following the setup in Section~\ref{sec:static_outdoor}. The victim car is traveling at $\approx$15~km/h, moving from $d_a$ = 20m to 3m distances from the van. The experiments are conducted under nighttime (50~lux) and daytime (850–1100~lux) conditions. We select patterns to induce false depth between 4-7m from the van. 

Figure~\ref{fig:van_experiment_setup} shows consistent incorrect depth estimation lasting more than 0.5~sec for both ZED2 and RealSense cameras under both lighting conditions. Specifically, false obstacles persist for 1.06~sec and 1.23~sec in daytime and nighttime, respectively, for RealSense and for 1~sec and 0.7~sec for ZED2. As in the case of the billboard pattern, the van is not in the victim’s trajectory; however, the attack succeeds in shifting the van's depth into the victim’s path along the frustum.
These findings confirm the possibility to conduct the attack by projecting onto real-world obstacles, with the potential to trigger unsafe and unexpected behaviors in the victim vehicles.

\section{High Speed Evaluation}
The attack under high‑speed driving conditions is evaluated on CARLA~\cite{dosovitskiy2017carla}, a widely used AV testing simulator. We synthesize a scenario in which the victim vehicle, equipped with a ZED2 stereo camera, drives at a constant speed, with a proxy billboard displaying the attack pattern placed 2m to the right of its trajectory, following the setup shown in Figure~\ref{fig:attack_scenario}(a). The vehicle approaches the billboard from 50m away and passes it at speeds of 10, 20, 30, and 40~km/h. We collect stereo images throughout the vehicle trajectory and apply the BM algorithm to estimate depth from the captured frames. $P_G$ and $P_{RS}$ are set to 26~pixels and 10~pixels, respectively, as they demonstrated the highest depth estimation error in static scenarios.

The results show consistent depth manipulation for durations of 5.3, 2.6, 2.3, and 2.1~s at vehicle speeds of 10, 20, 30, and 40~km/h, respectively. The shorter attack durations observed at higher speeds are because the vehicle traverses the fixed 50m distance in less time. These durations all exceed the 0.5~s chattering threshold, the minimum detection time required for the vehicle to react in autonomous driving frameworks such as Autoware~\cite{autoware}. These results indicate that the attack can induce consistent depth shifts at typical driving speeds of 40~km/h, potentially triggering unsafe driving behaviors, including sudden braking and unexpected maneuvers.
\section{Defenses}
\label{sec:defense}

While patterns with repeating elements are recognized as a challenging case for stereo depth estimation, stereo camera vendors recommend parameter adjustments to mitigate depth errors~\cite{realsense_fix}. For example, RealSense suggests changing the census register diameter and the peak threshold to account for depth errors.
Such tuning is impractical in dynamic real‑world environments where autonomous systems operate, because parameters would require continuous, real‑time adaptation to the spectral properties of incoming stereo images. Moreover, such tuning can only mitigate attacks at a single pattern granularity, whereas an adversary can vary across multiple granularities.

Barrois~et~al.~\cite{barrois2010resolving} propose detecting repetitive regions using FFT and 
to estimate the true distance using a stereo model‑free optimization. 
To verify the effectiveness of this method, we implement and evaluate it on 200 stereo image pairs collected from the ZED2 camera using the setup described in Section~\ref{sec:static_outdoor}, comprising daytime and nighttime scenarios. 
Since our patterns are placed at 3–20m away, we consider the attack mitigated when the depth estimation error is reduced to less than 0.5m, an acceptable relative error~\cite{llorca2010error}. 
As a result, the rate of successful mitigation is only 13\% of the evaluated stereo image regions. This lower success rate is primarily caused by the FFT‑based detector that fails to capture the repeated elements. As real‑world patterns do not remain perfectly uniform, this violates the assumptions in the original method, especially when $P_{RS}<P_G$. This is consistent with the design limitations discussed in the work, which require near-perfectly repeated structures.

\subsection{Proposed Defense}

To overcome the limitations of previous approaches, we propose a new defense that leverages the effect of repeated elements on matching scores (SAD) in BM algorithms. 
In the presence of repeated elements, the trace of the matching scores over the pixel offsets includes periodic peaks corresponding to the pattern, as shown in Figure~\ref{fig:sad_minima_variance}.
The proposed method uses such repeated peaks to detect patterns placed by an attacker.

To achieve this, we normalize the SAD scores for each block and identify minima produced by oscillations induced by repeated structures in the pattern. To ensure that these minima correspond to structured repetition rather than noise, we detect peaks only when they occur at regular intervals and exhibit consistent magnitude and gradient. For each block, the disparity range containing such detected oscillatory peaks is labeled as a repeated element, and applying this process across all blocks yields a mask of the attacked region. When identified, we draw a corresponding 2D bounding box and perform block matching using the entire bounding box as a single unit. Standard matching is then applied to the stereo image pair, and the minimum SAD score obtained for the bounding box is taken as the correct disparity estimation within the repeated pattern region. 
The proposed defense on 200 stereo image pairs collected under real‑world conditions achieves 96.5\% success rate in mitigating the attack, with the resulting post‑mitigation depth error remaining below 0.5m. This corresponds to a 98\% success rate in nighttime and 94.5\% in daytime conditions, demonstrating consistent performance across varying environments. 

We further extend our methodology to deep learning models. To achieve this, we extract the PSMNet model's intermediate results from the cost-volume matrix. The cost-volume matrix of deep learning-based SDE algorithms measures the feature-level disparity between stereo images, analogous to SAD scores in classical SDE algorithms. The feature magnitudes from the matrix are normalized, then equidistant minima are detected in the feature magnitudes, following the methodology described above for the SAD scores. To evaluate this, we apply the attack to the 200 image pairs from KITTI, using the same experimental setup described in Section~\ref{sec:experiment_setup}. This methodology achieves 100\% success rate in suppressing the depth estimation error to below 0.1m. These results demonstrate the effectiveness of the proposed defense in mitigating the attack against both classical and deep learning-based SDE algorithms.
Furthermore, the defense can also be used to mitigate natural repeated patterns, which can unintentionally trigger the same effects.
\section{Discussion}

\noindent\textbf{Training Dataset.}
Deep learning based models are affected by their training methodology. 
For example, when analyzing the UniMatch model pretrained on the SceneFlow dataset~\cite{MIFDB16} instead of KITTI, we observe a shift in the range of possible depth errors and the patterns that are most effective at a given distance. The maximum predicted disparity within the pattern region decreases by over 50 pixels at a true distance of 60 m, but increases by over 50 pixels at a true distance of 10 m, with a completely different selection of effective patterns. 
This shows how multiple factors concur to trigger the vulnerability in deep-learning models, including the dataset used, specific model architecture, and training methodology.
Additionally, convolutional aggregation in deep learning-based models can blend features extracted from the pattern and the background at the pattern edges, as shown in Figure~\ref{fig:edge_artifact}. Further details are in Appendix~\ref{appendix:global_dl}.

\noindent\textbf{Changing Illuminance and Texture.} The attacker can adjust the projection brightness depending on distance and environmental conditions, using the formula $B_p=B+k \cdot I$, where $B$ is the average projected pixel brightness at a distance, $k$ is the flat surface reflectance factor, and $I$ is the average ambient illumination. We use this during our daytime outdoor evaluation (natural light ranging from 1000-1500 lux) with a conventional projector at 4 meters.

Further, projection patterns can be adjusted based on the texture of the surface used for the projection. For example, on a surface with curvature $CU$, adversaries can use standard radial image contortions based on $\Delta i=i - (1/CU)\cdot \sin(i \cdot CU)$, where $i$ is the radial pixel distance, to maintain perspective across image frames.

\noindent\textbf{Stealthiness.} The adversary can further deploy a stealthy attack by disguising the repeated pattern with naturalistic textures by leveraging surrounding objects~\cite{zheng2024pi} or by exploiting the infrared domain, as demonstrated in prior work~\cite{sato2024invisible, bhupathiraju2024vulnerability} to project patterns. We additionally verify that adversaries can leverage infrared projections against the RealSense D435 camera, as some stereo systems do not deploy infrared filters to support time‑of‑flight depth estimation or to enhance low‑light performance.

\noindent\textbf{Limitations.} In our real‑world dynamic experiments, a single attack pattern is used to manipulate depth estimation on the ZED2 and RealSense cameras. 
A more sophisticated adversary, however, could employ dynamic patterns that adapt to the victim vehicle’s position, enabling longer‑lasting depth estimation errors while reducing the necessary attack surface. We further limit our experiments to patterns displayed or projected onto flat or semi-flat surfaces. In theory, adversaries could exploit obstacles with arbitrary shapes to project structured patterns, which make the precise obstacle depth estimation more challenging. Although our analysis is limited to four state-of-the-art stereo matching algorithms and deep learning models, current stereo processing techniques are derived from these approaches, which can potentially extend vulnerability risks to other architectures and stereo cameras within the same family.
\section{Related Works}
\label{sec:related_works}
Physical attacks for spoofing and modifying sensor readings are a growing concern. 
Prior works have demonstrated that cameras are susceptible to a range of spoofing attacks, including laser injection~\cite{nassi2019mobilbye, yan2016can, sato2024invisible, kohler2022signal} and electromagnetic interference attacks \cite{kohler2022signal}. Similarly, LiDARs and radars are also susceptible to such physical attacks~\cite{hunt2023madradar, cao2023you, bhupathiraju2023emi, jin2024phantomlidar, cao2019adversarial}.

These attacks have been shown to be highly effective within the ADAS domain, where sensor perturbations can affect downstream decision-making. Bhupathiraju et al. \cite{hrush2026thermal} demonstrate that thermal cameras are susceptible to attacks on their image processing by either naturally occurring or maliciously placed heat sources, causing failures in downstream object detection. Chen et al.~\cite{chen2024adversary} leverage repeating patterns and build optimized adversarial patterns to attack camera-based SLAM algorithms for monocular depth estimation. Nassi et al.~\cite{nassi2020phantom} exploit a vulnerability in object detectors that do not consider the depth of the obstacle. This work specifically considers stereo camera depth estimation, and the ambiguity caused by repeated elements within the scene. 

Deep Neural Networks also pose their own unique security concerns within the ADAS domain. In particular, prior works have demonstrated that Deep Neural Networks are highly susceptible to adversarial perturbations, including patch-based attacks \cite{wei2024physical} \cite{brown2017adversarial} \cite{eykholt2018robust}. In contrast, this work specifically uncovers a hidden vulnerability in stereo cameras that can be used to achieve controlled depth manipulation, without the need for adversarial optimization.

\section{Conclusion}
Our work identifies a dangerous vulnerability in stereo cameras that can be exploited using structured repeated patterns to induce depth estimation errors. These vulnerabilities also affect deep learning SDE algorithms without requiring the creation of adversarial examples. Our evaluation demonstrates that an adversary can take advantage of this vulnerability to control classical block matching algorithms and ML models, including two widely used stereo cameras deployed in autonomous systems. We evaluate attacks in high-speed simulated and controlled real‑world scenarios, showing their practicality in driving conditions. Finally, we design a defense method that detects repeated elements and mitigates the resulting depth estimation.

\section{Ethical Considerations}
\label{sec:ethics}
All experiments in this work use publicly available datasets and software. The use of publicly available datasets ensures transparency and reproducibility. The attack evaluation in our indoor and outdoor scenarios was conducted in controlled environments mimicking real‑world driving while adhering to the safety policy of our institutions and the speed limits, with the vehicles operating at up to 15~km/h. In line with ethical guidelines, we have disclosed our findings to the relevant stakeholders (ZED and RealSense) and are awaiting their responses. No human studies were involved in this research. 

\section{Acknowledgment}
This paper was edited for grammar using Grammarly and Microsoft Copilot. We disclosed the vulnerability and our findings to the vendor. This research was supported in part by the JST CREST JPMJCR23M4, JST FOREST JPMJFR2531, JST Next-generation Edge AI Semiconductor JPM-JES2515, and JSPS KAKENHI 24K02940 and 24K14943.

\bibliographystyle{ACM-Reference-Format}
\bibliography{bib}

\appendix
\section*{Appendix}

\section{Influence of Sampling Artifact on SAD Score}
\label{appendix:sad_influence}

Here, we formalize how sampling artifacts affect pixel intensities and, in turn, bias the minima of the block matching SAD cost, shifting predicted disparity. For a physical point $p=(x,y,z)$, the pixel intensity can be estimated using Equation~\ref{eq:pixel_intensity1}. The corresponding pixel intensity $P^R$ of the $i^{th}$ pixel in the right image is given by:
\begin{equation}
P^R(i+D) = \frac{1}{A(SR_i)} \int_{SR_{i+D}}^{SR_{i+D+1}} I_\mathrm{opt}(p) \mathrm{d}A \hspace{1em} 
\label{eq:pixel_intensity2}
\end{equation}

Here, $D$ denotes the ground truth pixel disparity between left and right images, computed as $D = \frac{f \cdot b}{d_a}$, where f is the focal length and b is the baseline distance between the two cameras (12~cm for ZED2). 

To model the variations, we first consider a simple black and white stripe pattern, placed at a distance $d_a$ from a stereo camera. We assume that each stripe in the pattern has a width $d_w$. 
The pixel intensity of the sampling artifact varies based on the optical intensity received from point $p$ in the repeated pattern, defined as: 

\begin{equation}
I_\mathrm{opt}(p) =
\begin{cases}
    1 & \text{if } 0 \le (p(x) \bmod d_w) < d_w/2 \\
    0 & \text{Otherwise}
\end{cases}
\label{eq:sampling_trend}
\end{equation}

Here, $p(x)$ defines the x-coordinate of the physical point $p$. This equation models pixel‑level inconsistencies in sampling artifacts that ultimately introduce measurable disparity offsets. These offsets arise from variations in the sampling artifacts, which differ in magnitude between the left and right images. Such subtle differences in the sampling artifacts ultimately influence the final matching in SDE algorithms.

\noindent\textbf{Influence on SAD Score.}
In BM algorithms, a block $B_L$ in the left image is compared with the a candidate block $B_R$ on the right image within the disparity search range. 
The algorithm finally selects a corresponding block $B_D$ in the right image that minimizes the SAD score, as given by 
\begin{equation}
B_D = \arg \min_{B_R} (\text{SAD}(B_L, B_R)) \hspace{1em} .
\label{eq:right_matching}
\end{equation}
When repeated elements are in the scene, multiple blocks in the right image ($B_R$) have similar small SAD scores due to their inherent similarity. This creates an oscillating trend in the estimated SAD scores, as illustrated in Figure~\ref{fig:sad_minima_variance} (blue oscillating pattern). Without sampling artifacts, these blocks demonstrate the same SAD score at the minima of the oscillating pattern, as shown in Figure~\ref{fig:sad_minima_variance} (a).
In such cases, the block with the highest disparity (i.e., the closest obstacle) is typically selected. In reality, however, each $B_R$ has a different SAD due to sampling artifacts: a particular $B_R$ with an artifact similar to that of $B_L$ has a lower SAD score than the others. The relative position between a pixel and the repeated pattern determines the pixel intensity,
as suggested by Equation~\ref{eq:pixel_intensity1}, and the SAD scores over a disparity become periodically high and low due to different intervals of pixels and the repeated pattern, as illustrated in Figure~\ref{fig:sad_minima_variance} (b). Figure~\ref{fig:sad_minima_variance} annotates the difference in the minima of the SAD scores with and without the presence of the SAD score, illustrating the influence on the minima of the oscillations. As we discuss further, this minimum ultimately influences the matching produced by the BM algorithm.


To further analyze the influence of sampling artifacts on disparity matching, we formalize the vulnerability.
We define an image block at position $i=(i_x,i_y)$ in the image as $B(i)$. Since block matching in stereo vision is performed along the horizontal axis, we focus on the $x$-direction. Let $B(i_\mathrm{Ref})$ denote the reference block in the left image, $B(i_T)$ the true corresponding block in the right image, and $B(i_F)$ a falsely matched block in the right image. The SAD score $B(i_\mathrm{Ref})$ and an arbitrary block $B(i)$ in the right image is expressed as $SAD(B(i_\mathrm{Ref}),B(i))$.

As shown in Figure~\ref{fig:sad_minima_variance} (left), the SAD scores become periodically high and low, reflecting the width $d_w$ of the repeating element width, which is expressed as
\begin{equation}
\mathrm{SAD}(B(i_\mathrm{Ref}), B(i)) \approx \mathrm{SAD}(B(i_\mathrm{Ref}), B(i_\mathrm{T}+n \cdot d_w)),
\label{eq:sampling_matching}
\end{equation}
where $T+n \cdot d_w$ represents a horizontal shift by $n$ repetitions of the pattern, with $n \in \{1, 2, \cdots\}$. This shows that the SAD score for $B(i_\mathrm{Ref})$ is approximately the same at each repeated element in the right image, since the pattern reappears every $d_w$ pixels. Such periodicity produces multiple candidate regions with similar SAD values, each of which can lead to a false match. This behavior is typical of block matching algorithms, which rely on local pixel intensities.

We define the average pixel intensity of the sampling artifact at contrasting boundaries (Figure~\ref{fig:stereo_causalities} (a)) as the sampling artifact magnitude. Let the magnitude at block position $i=(i(x),i(y))$ be denoted as $P(i)$, where $i(x)=n\cdot d_w$ (at every repetition of the pattern). Here, $P(i)$ is estimated from the pixel intensity using Equation~\ref{eq:pixel_intensity2}. The SAD score depends on the similarity between the sampling artifacts of two blocks:
\begin{equation}
\mathrm{SAD}(B(i_\mathrm{Ref}), B(i)) \propto |P(i_\mathrm{Ref}) - P(i)|.
\end{equation}

Thus, for repeating patterns, the SAD score between a reference block and a potential match ($i(x)=i_T(x)+n\cdot dw$) is proportional to the difference in sampling artifact magnitude. An incorrect match occurs when
\begin{equation}
\mathrm{SAD}(B(i_\mathrm{Ref}), B(i_{F})) < \mathrm{SAD}(B(i_\mathrm{Ref}), B(i_{T}))),
\end{equation}
which implies that the sampling artifact magnitude of the reference block is closer to that of the false match than to the true match. This condition can be expressed as:
\begin{equation}
|P(i_\mathrm{Ref}) - P(i_\mathrm{F})| < |P(i_\mathrm{Ref}) - P(i_\mathrm{T})|.
\end{equation}

The minima of the SAD scores, which determine the optimal match in BM algorithms, become dominated by sampling artifacts (Figure~\ref{fig:sad_minima_variance}). Consequently, the BM algorithm tends to select the offset $n$ that minimizes the difference:

\begin{equation}
\arg\min_{n} \left| P(i_\mathrm{Ref}) - P\!\left(i_{T} + n \cdot d_w\right) \right|.
\label{eq:artifact_solving}
\end{equation}

This leads to a lower matching cost but an incorrect disparity estimate. Because the sampling artifact magnitude $P(i)$ varies periodically with $d_w$, as described in Equation~\ref{eq:sampling_trend}, the minima of the SAD score are directly influenced. Since disparity estimation relies on these minima, the sampling artifact ultimately affects which block is chosen as the correct match when repeated elements create multiple candidates. An adversary could exploit this by estimating the periodic trend of sampling artifact intensity from Equation~\ref{eq:sampling_trend} and adjusting the pattern according to Equation~\ref{eq:artifact_solving} to manipulate depth estimation.

\section{Sampling Artifact Consistency in PSMNet}
\label{appendix:sampling_consistency}
\begin{figure}[t]
    \centering
   \includegraphics[width=\linewidth]{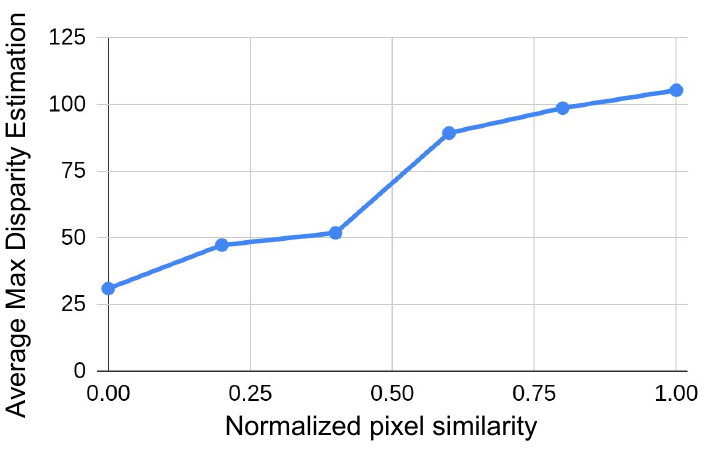}%
    \caption{The average maximum disparity increases with respect to higher normalized pixel similarity (higher similarity in sampling artifact pixel). The graph further validates that feature matching in deep learning-based SDE models is influenced by the sampling artifacts.}
    \Description{Avg. Max. Disp. Increase.}\label{fig:sampling_further_PSMNet}
\end{figure}

To further validate the effect of sampling artifacts on the PSMNet model, we compute the pixel similarity between the left and right images within the artifact region using normalized pixel similarity, defined as the absolute difference between the sampling artifact intensities in each image pair. A high normalized similarity indicates that the corresponding artifact magnitudes are closely matched. The maximum disparity is measured with increments of 0.2, corresponding to these similarity values. As shown in Figure~\ref{fig:sampling_further_PSMNet}, artifacts with higher normalized similarities (similar sampling artifact magnitudes) are consistently matched, mirroring the behavior of classical block-matching algorithms. This further confirms that the vulnerability observed in BM algorithms also extends to deep learning models such as PSMNet.

\begin{figure}[t]
    \centering
   \includegraphics[width=\linewidth]{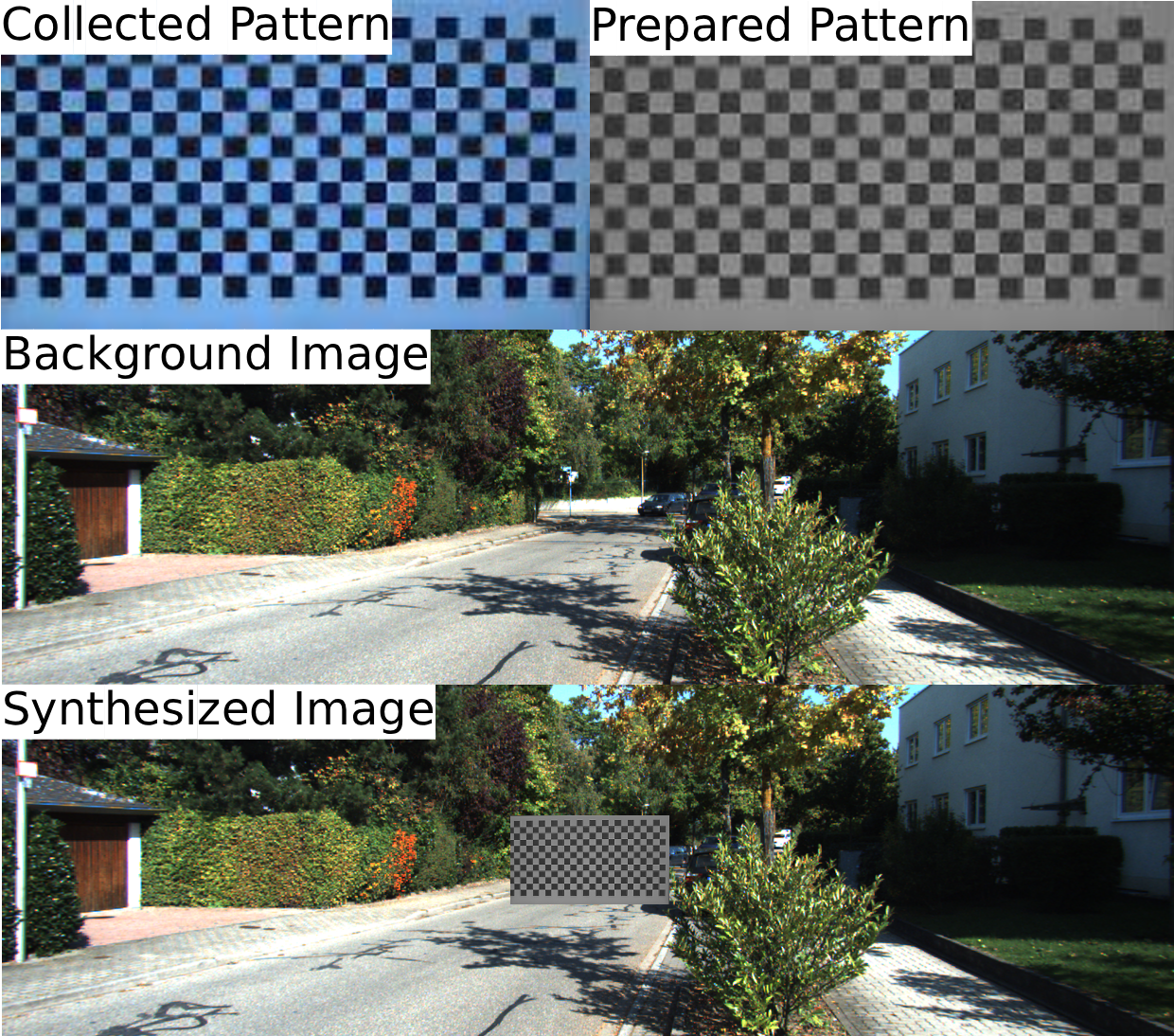}%
    \caption{Image synthesis with an attack pattern. Each pattern is cropped from a real-world image of a computer monitor, then luminance matched to the background image, and placed on top of the image with the corresponding disparity.} 
    \Description{Attack pattern synthesis}
    \label{fig:kitti_synthesis}
\end{figure}





\section{Global Properties of Deep-learning Models.}
\label{appendix:global_dl}
\begin{figure}[t]
    \centering
   \includegraphics[width=\linewidth]{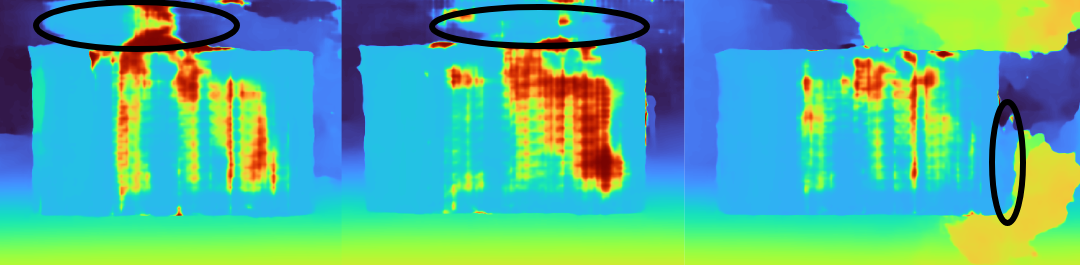}%
    \caption{Edge artifacts caused by the pattern on three different scenes.}
    \Description{Edge artifacts caused by the pattern on three different scenes.}
    \label{fig:edge_artifact}
\end{figure}

Deep learning–based SDE models typically rely on convolutional layers to estimate disparity by analyzing local similarity between the left and right images. Consequently, at the boundaries of the attack pattern, features from both the pattern and the surrounding foreground are blended through convolutional aggregation. This interaction can create a perspective hijack, leading to incorrect depth estimation for objects located at the edges of the attack region. Examples of such edge‑based errors are shown in Figure~\ref{fig:edge_artifact}.

We do not include these edge‑induced disparity errors in our evaluation, as they arise specifically from the convolutional feature extraction of deep learning models. Nonetheless, adversaries could exploit this behavior to mislead SDE models and extend the affected region of a scene.

\section{Motion Blur Simulation}
\label{appendix:motionblur}
The motion-blurred image can be modeled by integrating the continuously changing image during the exposure interval \(T\). If the camera moves toward the obstacle with constant velocity \(v\), the blurred image is given by

\begin{equation}
I_b(x,y) = \frac{1}{T}\int_0^T I\!\left(x-\Delta x(t),\, y-\Delta y(t)\right) dt,
\end{equation}

where \((\Delta x(t), \Delta y(t))\) represents the image displacement caused by camera motion at time \(t\). For forward motion, the displacement is radial and increases with distance from the image center:

\begin{equation}
\Delta r(t) = \frac{fvt}{Z(x,y)},
\end{equation}

where \(f\) is the focal length and \(Z(x,y)\) is the scene depth. Thus, each pixel accumulates intensity along a radial trajectory during exposure, producing a spatially varying radial motion blur. The equivalent point spread function corresponds to the normalized trajectory traced by the pixel over the exposure duration \(T\).

\section{Open Science Policy}
We are dedicated to promoting the principles of open science and ensuring the reproducibility of research in the field of autonomous system security. To uphold this commitment, we provide comprehensive access to all relevant research artifacts
and materials as detailed below.
Details and demo videos recordings are available on our website: \url{https://sites.google.com/view/stereocam/home}. Our research artifacts, including the scripts to synthesize and evaluate the attack on SDE models, the resulting depth estimation point clouds from real-world experiments, and the script to test and evaluate the defense methodology, are available at: \url{https://zenodo.org/records/20767261}.

\end{document}